\documentclass[a4paper,12pt]{article}

\pdfoutput=1

\usepackage{amsmath}
\usepackage{amssymb}
\usepackage{mathrsfs}
\usepackage{bbm}
\usepackage{graphicx,subfigure}
\usepackage[numbers,sort&compress]{natbib}
\usepackage{array,longtable}
\usepackage{multirow}

\usepackage[utf8]{inputenc}

\usepackage[colorlinks,
            linkcolor=black,
            filecolor=black,
            anchorcolor=black,
            urlcolor=black,
            citecolor=blue,
            bookmarks=false,
            ]{hyperref}

\numberwithin{equation}{section}

\def\dd{{\rm d}}
\def\mLambdap{(m_{\Lambda_b}+m_{\Lambda_c})}
\def\mLambdam{(m_{\Lambda_b}-m_{\Lambda_c})}

\newlength{\dinwidth}
\newlength{\dinmargin}
\allowdisplaybreaks

\begin{document}

\title{\bf $\boldsymbol{b\to c\tau\nu}$ Transitions in the \\[1mm] Standard Model Effective Field Theory}

\author{
Quan-Yi Hu\footnote{qyhu@mails.ccnu.edu.cn},
Xin-Qiang Li\footnote{xqli@mail.ccnu.edu.cn}\,
and
Ya-Dong Yang\footnote{yangyd@mail.ccnu.edu.cn}\\[15pt]
\small Institute of Particle Physics and Key Laboratory of Quark and Lepton Physics~(MOE), \\
\small Central China Normal University, Wuhan, Hubei 430079, China}

\date{}
\maketitle

\vspace{-0.2cm}

\begin{abstract}
{\noindent}The $R(D^{(\ast)})$ anomalies observed in $B\to D^{(\ast)}\tau\nu$ decays have attracted much attention in recent years. In this paper, we study the $B\to D^{(\ast)}\tau\nu$, $\Lambda_b\to\Lambda_c\tau\nu$, $B_c\to (J/\psi,\,\eta_c)\tau\nu$, $B\to X_c\tau\nu$, and $B_c\to\tau\nu$ decays, all being mediated by the same quark-level $b\to c\tau\nu$ transition, in the Standard Model Effective Field Theory. The most relevant dimension-six operators for these processes are $Q_{lq}^{(3)}$, $Q_{ledq}$, $Q^{(1)}_{lequ}$, and $Q^{(3)}_{lequ}$ in the Warsaw basis. Evolution of the corresponding Wilson coefficients from the new physics scale $\Lambda=1$~TeV down to the characteristic scale $\mu_b\simeq m_b$ is performed at three-loop in QCD and one-loop in EW/QED. It is found that, after taking into account the constraint ${\cal B}(B_c\to\tau\nu)\lesssim 10\%$, a single $\left[C_{lq}^{(3)}\right]_{3323}(\Lambda)$ or $\left[C^{(3)}_{lequ}\right]_{3332}(\Lambda)$ can still be used to resolve the $R(D^{(\ast)})$ anomalies at $1\sigma$, while a single $\left[C^{(1)}_{lequ}\right]_{3332}(\Lambda)$ is already ruled out by the measured $R(D^{(\ast)})$ at more than $3\sigma$. By minimizing the $\chi^2(C_i)$ function constructed based on the current data on $R(D)$, $R(D^\ast)$, $P_\tau(D^\ast)$, $R(J/\psi)$, and $R(X_c)$, we obtain eleven most trustworthy scenarios, each of which can provide a good explanation of the $R(D^{(\ast)})$ anomalies at $1\sigma$. To further discriminate these different scenarios, we predict thirty-one observables associated with the processes considered under each NP scenario. It is found that most of the scenarios can be differentiated from each other by using these observables and their correlations.
\end{abstract}

\newpage

\section{Introduction}
\label{sec:introduction}

In the past few years, the $B$-physics experiments have reported a number of interesting anomalies in the semi-leptonic $B$-meson decays, which have aroused a lot of attention~\cite{Bifani:2018zmi,Li:2018lxi,Ciezarek:2017yzh}. In the charged-current processes $B\to D^{(\ast)}\ell\nu$, for example, the ratios of the branching fractions\footnote{The advantage of considering the ratios $R(D^{(\ast)})$ instead of the branching fractions themselves lies in the fact that, apart from the significant reduction of the experimental systematic uncertainties, the CKM matrix element $V_{cb}$ cancels out and the sensitivity to the $B\to D^{(\ast)}$ transition form factors becomes much weaker.}
\begin{equation}
R(D^{(\ast)})=\frac{{\cal B}(B\to D^{(\ast)}\tau\nu)}{{\cal B}(B\to D^{(\ast)}\ell\nu)},
\end{equation}
have been measured by the BaBar~\cite{Lees:2012xj,Lees:2013uzd} and Belle~\cite{Huschle:2015rga,Sato:2016svk,Hirose:2016wfn,Hirose:2017dxl} collaborations with $\ell=e,\,\mu$, as well as the LHCb~\cite{Aaij:2015yra,Aaij:2017uff,Aaij:2017deq} experiment with $\ell=\mu$. These measurements have been averaged by the Heavy Flavor Averaging Group (HFLAV)~\cite{Amhis:2016xyh}, and the latest results read~\cite{Amhis:2018up}
\begin{equation}\label{eq:rddstar_exp}
R(D)^{\text{avg}}=0.407 \pm 0.039 \pm 0.024,\quad
R(D^\ast)^{\text{avg}}=0.306 \pm 0.013 \pm 0.007,
\end{equation}
with a correlation of $-0.203$. Comparing Eq.~\eqref{eq:rddstar_exp} with the arithmetic average~\cite{Amhis:2018up} of the latest Standard Model (SM) predictions~\cite{Bigi:2016mdz,Bernlochner:2017jka,Bigi:2017jbd,Jaiswal:2017rve},
\begin{equation}\label{eq:rddstar_theo}
R(D)=0.299 \pm 0.003,\quad R(D^\ast)=0.258 \pm 0.005,
\end{equation}
one can see that the difference between experiment and theory is at about $3.78\sigma$ corresponding to $99.98\%$ confidence level (C.L.), implying therefore intriguing hints of lepton-flavour universality violating new physics (NP) beyond the SM. To understand these anomalies, many studies have been done; see for instance Ref.~\cite{Hu:2018lmk} and references therein, as well as Refs.~\cite{Carena:2018cow,Cohen:2018vhw,Heeck:2018ntp,Angelescu:2018tyl,Huang:2018nnq,Azatov:2018kzb,Li:2018rax,Robinson:2018gza,Trifinopoulos:2018rna,Feruglio:2018fxo,Kumar:2018kmr,Becirevic:2018afm,Dhargyal:2018bbc,Bhattacharya:2018kig,Fraser:2018aqj,Martinez:2018ynq,Azatov:2018knx,Alok:2018uft,Aydemir:2018cbb,Greljo:2018ogz,Asadi:2018wea,Marzocca:2018wcf,Yang:2018pyq,Iguro:2018qzf,Colangelo:2018cnj,Biswas:2018jun,Jung:2018lfu,Choudhury:2017ijp,He:2017bft,Bardhan:2016uhr,Crivellin:2015hha,Fajfer:2012vx,Fajfer:2012jt,Crivellin:2012ye,Alok:2017qsi}.

On the other hand, in view of the absence (so far) of any clear signal of new particles at the LHC, the NP scale $\Lambda$ should be much higher than the electroweak (EW) scale $\mu_\text{EW}\simeq 246$~GeV. Assuming further that there exist no undiscovered but weakly coupled light particles, any NP effect in the processes proceeding at energy scales well below $\Lambda$ but above $\mu_\text{EW}$ can be effectively described by a series of higher dimensional operators that are built out of the SM fields and are invariant under the SM gauge group $SU(3)_C\otimes SU(2)_L\otimes U(1)_Y$~\cite{Buchmuller:1985jz,Grzadkowski:2010es}. The resulting effective field theory (EFT) is conventionally called the Standard Model Effective Field Theory (SMEFT)~\cite{Weinberg:1980wa,Coleman:1969sm,Callan:1969sn}, which has now emerged as one of the most interesting tools to probe systematically the data from the LHC and elsewhere for possible NP hints\footnote{See, for example, Refs.~\cite{Willenbrock:2014bja,Falkowski:2015fla,Brivio:2017vri,deFlorian:2016spz} for recent reviews on the SMEFT.}. For energies below $\Lambda$, the leading NP contributions in the SMEFT formalism arise from the dimension-six operators\footnote{There exists only a single dimension-five operator in the SMEFT, up to Hermitian conjugation and flavour assignments~\cite{Weinberg:1979sa,Grzadkowski:2010es}. It violates the lepton number and, after the EW symmetry breaking, gives Majorana masses for the SM neutrinos. This operator is irrelevant to this paper.}, which were firstly classified in Ref.~\cite{Buchmuller:1985jz}, but found to be redundant for some of them. The first complete and non-redundant basis of dimension-six operators was derived in Ref.~\cite{Grzadkowski:2010es} and is now commonly called the Warsaw basis\footnote{Apart from the Warsaw basis~\cite{Grzadkowski:2010es}, other bases were also proposed, with the most prominent ones being the HISZ~\cite{Hagiwara:1993ck} and the SILH~\cite{Contino:2013kra,Giudice:2007fh} basis. For an easy translation between these different bases, one can resort to the computer codes \texttt{Rosetta}~\cite{Falkowski:2015wza} and \texttt{WCxf}~\cite{Aebischer:2017ugx}.}. The complete one-loop anomalous dimensions of these dimension-six operators have also been calculated in Refs.~\cite{Jenkins:2013zja,Jenkins:2013wua,Alonso:2013hga}.

The EFT approach is also an essential ingredient for $B$-physics analyses both within and beyond the SM. As the typical energy scale $\mu_b$ is around the bottom-quark mass $m_b\simeq5$~GeV, being much smaller than the EW and the NP scale, all the $B$-physics processes can be well described by an effective Lagrangian constructed by integrating out the SM and NP heavy degrees of freedom (for classical reviews, see for example Refs.~\cite{Buchalla:1995vs,Buras:1998raa}). The resulting EFT includes only the QCD and QED gauge interactions coupled to all the six leptons and the five lightest quarks, plus a full set of dimension-six local operators built with these matter fields as well as the gluon and photon field-strength tensors, and is conventionally called the weak effective theory (WET)~\cite{Aebischer:2017gaw,Jenkins:2017jig,Jenkins:2017dyc}. In contrast to the SMEFT case, the dimension-six operators in WET are not invariant under the full SM gauge group, but only under $SU(3)_C\otimes U(1)_{\rm em}$, as this EFT is defined below the EW scale where $SU(2)_L\otimes U(1)_Y$ is already broken. A complete and non-redundant set of dimension-six operators relevant for $B$ physics, together with the complete one-loop anomalous dimensions in QCD and QED, can be found in Refs.~\cite{Aebischer:2017gaw,Jenkins:2017jig,Jenkins:2017dyc}.

For a given set of SMEFT dimension-six operators with the corresponding Wilson coefficients specified at the scale $\Lambda$, to study their effects on the $B$-physics processes, one has to follow the following three steps~\cite{Aebischer:2018bkb}: perform the renormalization group evolution (RGE) of the SMEFT Wilson coefficients from the NP down to the EW scale~\cite{Jenkins:2013zja,Jenkins:2013wua,Alonso:2013hga}; match the given set of SMEFT operators onto the WET ones at the EW scale~\cite{Jenkins:2017jig,Aebischer:2015fzz}; perform the RGE of the WET Wilson coefficients from the EW down to the scale $\mu_b$~\cite{Aebischer:2017gaw,Jenkins:2017jig,Jenkins:2017dyc}. With the aid of these three steps, one can then bridge the gap between the SMEFT Lagrangian and the low-energy measurements in $B$ physics. In this paper, following this procedure and motivated by the $R(D^{(\ast)})$ anomalies, we shall study the $B\to D^{(\ast)}\tau\nu$, $\Lambda_b\to\Lambda_c\tau\nu$, $B_c\to (J/\psi,\,\eta_c)\tau\nu$, $B\to X_c\tau\nu$, as well as $B_c\to\tau\nu$ decays, all being mediated by the same quark-level $b\to c\tau\nu$ transition, in the SMEFT formalism. It is found that the most relevant operators for these processes are $Q_{lq}^{(3)}$, $Q_{ledq}$, $Q^{(1)}_{lequ}$, and $Q^{(3)}_{lequ}$ in the Warsaw basis. The RGEs of the corresponding Wilson coefficients from the NP scale $\Lambda$ down to the typical scale $\mu_b$ is performed at three-loop in QCD and one-loop in EW/QED (see Refs.~\cite{Feruglio:2016gvd,Feruglio:2017rjo,Gonzalez-Alonso:2017iyc} and references therein). Confronted with the currently available data, we shall also perform a detailed phenomenological analysis of these decays.

Our paper is organized as follows. In section \ref{sec:Lagrangian}, after recapitulating the SMEFT Lagrangian, we list the most relevant dimension-six operators for $b\to c\tau\nu$ transitions, and then discuss the evolution and matching of these operators in both the SMEFT and WET. In section \ref{sec:observables}, all the observables considered in the paper are listed, and the corresponding inputs for the transition form factors are also mentioned. Our numerical results and discussions are presented in section \ref{sec:results}. Finally, we make our conclusions in section \ref{sec:conclusion}.  Explicit expressions of the helicity amplitudes for $\Lambda_b\to \Lambda_c\tau\nu$ decay are collected in the appendix.

\section{Theoretical framework}
\label{sec:Lagrangian}

\subsection{SMEFT Lagrangian}

Following the common practice to truncate the SMEFT Lagrangian at dimension-six level and assuming that the EW symmetry breaking is realized linearly, we can write the SMEFT Lagrangian as
\begin{equation}\label{eq:smeft}
  \mathcal{L}_\mathrm{SMEFT}=\mathcal{L}_\mathrm{SM}^{(4)}+\frac{1}{\Lambda^2}\sum_{i}C_i(\Lambda)Q_i\,,
\end{equation}
where $\mathcal{L}_\text{SM}^{(4)}$ is the usual SM Lagrangian before spontaneous symmetry breaking (SSB). The dimension-six operators $Q_i$, which are obtained by integrating out all the heavy NP particles and are invariant under the SM gauge symmetry, are given by
\begin{align}\label{eq:SMEFT_O}
Q_{lq}^{(3)}=\:&(\bar{l}\gamma_\mu\tau^I l)(\bar{q}\gamma^\mu\tau^I q),&
Q_{ledq}=\:&(\bar{l}^je)(\bar{d}q^j),\nonumber\\[2mm]
Q^{(1)}_{lequ}=\:&(\bar{l}^je)\varepsilon_{jk}(\bar{q}^ku),&
Q^{(3)}_{lequ}=\:&(\bar{l}^j\sigma_{\mu\nu}e)\varepsilon_{jk}(\bar{q}^k\sigma^{\mu\nu}u),
\end{align}
and so on~\cite{Buchmuller:1985jz,Grzadkowski:2010es}. Here $\tau^I$ are the Pauli matrices, and $\varepsilon_{jk}$ is the totally antisymmetric tensor with $\varepsilon_{12}=+1$. The fields $q$ and $l$ correspond to the quark and lepton $SU(2)_L$ doublets, while $u$, $d$ and $e$ are the right-handed $SU(2)_L$ singlets. All the NP contributions are encoded in the Wilson coefficients $C_i$, which are dependent on the renormalization scale. This scale dependence will, however, be canceled in a physical amplitude by that of the matrix elements of $Q_i$.

In this paper, we focus only on the operators $Q_{lq}^{(3)}$, $Q_{ledq}$, $Q^{(1)}_{lequ}$ and $Q^{(3)}_{lequ}$, as well as their hermitian conjugates, which contribute to the $b\to c\tau\nu$ transitions at tree level~\cite{Gonzalez-Alonso:2017iyc,Aebischer:2015fzz}. Note that the operator $Q_{lq}^{(3)}$ is already self-conjugate~\cite{Buchmuller:1985jz,Grzadkowski:2010es}. We also assume that the flavour of the neutrino in these operators is pure $\nu_\tau$.

\subsection{Evolution and matching}

To explore the NP effect on the $b\to c\tau\nu$ transitions, we should firstly link the SMEFT Lagrangian given at the NP scale $\Lambda$ to the WET Lagrangian given at the typical energy scale $\mu_b$ associated with the processes considered. This can be achieved through the following three steps, details of which could be found, for example, in Refs.~\cite{Aebischer:2015fzz,Aebischer:2017gaw,Jenkins:2017jig,Jenkins:2017dyc,Aebischer:2018bkb}.

Firstly, we should evolve the Wilson coefficients $C_i$ of the SMEFT Lagrangian from the initial scale $\Lambda$ down to the EW scale $\mu_\text{EW}$, under the SM gauge group $SU(3)_C\otimes SU(2)_L\otimes U(1)_Y$. For simplicity, here we do not discriminate the masses of $W^\pm$, $Z^0$, the top quark $t$, and the Higgs boson $h$, and set approximately all of them to be $\mu_\text{EW}$. The one-loop RGE flow of $C_i(\mu)$ can be written schematically as
\begin{equation}\label{eq:RGEhigh}
\mu\,\frac{\dd C_i}{\dd\mu}=\frac{1}{16\pi^2}\sum_j\gamma_{ij}C_j\equiv\frac{1}{16\pi^2}\beta_i\,.
\end{equation}
Neglecting terms suppressed by the Yukawa couplings, which are found to be negligibly small in our case, the one-loop beta functions are given, respectively, by~\cite{Jenkins:2013zja,Jenkins:2013wua,Alonso:2013hga,RGEweb}
\begin{align}
\left[\beta_{lq}^{(3)}\right]_{prst} & = \frac{2}{3}g^2\left\{3\left[ C_{lq}^{(3)}\right]_{prww}\delta_{st}+\left[ C_{lq}^{(3)}\right]_{wwst}\delta_{pr}\right\}-(6g^2+g'^2)\left[ C_{lq}^{(3)}\right]_{prst},\\[2mm]
\beta_{ledq} & = -\left(\frac{8}{3}g'^2+8g_s^2\right)C_{ledq},\\[2mm]
\beta_{lequ}^{(1)} & = -\left(\frac{11}{3}g'^2 + 8 g_s^2\right) C_{lequ}^{(1)}+\left(30g'^2 + 18 g^2\right) C_{lequ}^{(3)},\\[2mm]
\beta_{lequ}^{(3)} & = \left(\frac{2}{9}g'^2 - 3g^2 +\frac{8}{3}g_s^2\right) C_{lequ}^{(3)}+\frac{1}{8}\left(5g'^2 + 3 g^2\right) C_{lequ}^{(1)}.\label{eq:betahigh}
\end{align}
Here we have introduced the abbreviations $\left[ C_{lq}^{(3)}\right]_{\cdot ww\cdot}\equiv \sum_w \left[ C_{lq}^{(3)}\right]_{\cdot ww\cdot}$, with $p,r,s,t,w$ being the flavour indices of the fermion fields in the weak-eigenstate basis, and $g_s$, $g$ and $g^{\prime}$ are the $SU(3)_C$, $SU(2)_L$ and $U(1)_Y$ gauge couplings, respectively. The SMEFT Lagrangian will undergo the SSB at an energy scale close to $\mu_\text{EW}$, making it necessary to switch from the weak to the mass eigenstates for the fermions. Performing the same flavour transformations as in Refs.~\cite{Aebischer:2015fzz,Jenkins:2017jig,Jenkins:2017dyc,Dedes:2017zog}, we can write the spontaneously broken SMEFT Lagrangian in terms of the mass-eigenstate fermion fields ($f_{L,R}^\text{(weak)}=P_{L,R}f^\text{(mass)}$) except the left-handed $d$-type quarks, for which the usual relation between the weak and mass eigenstates reads~\cite{Jenkins:2017jig}
\begin{equation}\label{eq:CKMeff}
d_{Lm}^\text{(weak)}=V_{md}P_L d^\text{(mass)}+V_{ms}P_L s^\text{(mass)}+V_{mb}P_L b^\text{(mass)}\equiv \sum_n V_{mn}P_L d_{n}^\text{(mass)}\,,
\end{equation}
where $P_{R,L}\equiv \frac{1\pm\gamma_5}{2}$ are the right- and left-handed chiral projectors. As we are concerned mainly on the operators $Q_{lq}^{(3)}$, $Q_{ledq}$, $Q^{(1)}_{lequ}$ and $Q^{(3)}_{lequ}$, as well as their hermitian conjugates, the effective quark-mixing matrix $V$ appearing in Eq.~\eqref{eq:CKMeff} coincides with the SM CKM matrix.

The second step is to perform the matching at the EW scale $\mu_\text{EW}$. After integrating out the SM heavy particles, the $W^\pm$, $Z^0$, the top quark, and the Higgs boson, we can obtain the WET Lagrangian suitable for describing the $b\to c\tau\nu$ transitions~\cite{Aebischer:2015fzz,Aebischer:2017gaw,Jenkins:2017jig}
\begin{equation}\label{eq:LangWET}
\mathcal{L}_\text{WET}=\mathcal{L}_\text{QCD+QED}^{(u,d,c,s,b,e,\mu,\tau,\nu_e,\nu_\mu,\nu_\tau)}+\mathcal{L}_\text{SM}^{(6)}+\mathcal{L}_\text{NP}^{(6)},
\end{equation}
where $\mathcal{L}_\text{QCD+QED}^{(u,d,c,s,b,e,\mu,\tau,\nu_e,\nu_\mu,\nu_\tau)}$ is the QCD and QED Lagrangian with all the six leptons and the five lightest quarks as the active degrees of freedom for fermions, and
\begin{align}\label{eq:WETSM}
\mathcal{L}_\text{SM}^{(6)} & =-\frac{4G_F}{\sqrt{2}}\,V_{cb}\,{\cal O}_{V_L}+\text{h.c.},\\[2mm]
\label{eq:WETNP}
\mathcal{L}_\text{NP}^{(6)} & = -\frac{4G_F}{\sqrt{2}}\,V_{cb}\,\big( C_{V_L}{\cal O}_{V_L}+C_{V_R}{\cal O}_{V_R}+C_{S_L}{\cal O}_{S_L}+C_{S_R}{\cal O}_{S_R}+C_T{\cal O}_T\big)+\text{h.c.},
\end{align}
with the WET dimension-six operators given, respectively, by\footnote{Neutrinos are assumed to be left-handed throughout this paper and, hence, we need not consider the tensor operator $({\bar c}\sigma^{\mu\nu} P_R b)({\bar\tau}\sigma_{\mu\nu} P_L\nu)$, which is obtained from ${\cal O}_T$ by changing the chirality of the quark current, because it is identically zero due to Fierz transformations.}
\begin{align}\label{eq:WET_O}
{\cal O}_{V_{L(R)}} & =({\bar c}\gamma^\mu P_{L(R)} b)({\bar\tau}\gamma_\mu P_L\nu),\nonumber\\[2mm]
{\cal O}_{S_{L(R)}} & =({\bar c}P_{L(R)} b)({\bar\tau}P_L\nu),\nonumber\\[2mm]
{\cal O}_T & =({\bar c}\sigma^{\mu\nu} P_L b)({\bar\tau}\sigma_{\mu\nu} P_L\nu).
\end{align}
Matching at tree level the SMEFT operators given by Eq.~\eqref{eq:SMEFT_O} onto the WET ones given by Eq.~\eqref{eq:WET_O} at the scale $\mu_\text{EW}$, we get~\cite{Aebischer:2015fzz,Jenkins:2017jig}
\begin{align}
C_{V_L} & =-\frac{\sqrt{2}}{2G_F\Lambda^2}\sum_{n}\left[ C_{lq}^{(3)}\right]_{332n}\frac{V_{nb}}{V_{cb}},
&
C_{S_R} & =-\frac{\sqrt{2}}{4G_F\Lambda^2}\frac{1}{V_{cb}}\left[ C_{ledq}\right]_{3332}^\ast,\nonumber
\\[2mm]
C_{S_L} & =-\frac{\sqrt{2}}{4G_F\Lambda^2}\sum_{n}\left[C^{(1)}_{lequ}\right]^\ast_{33n2}\frac{V_{nb}}{V_{cb}},
&
C_T & =-\frac{\sqrt{2}}{4G_F\Lambda^2}\sum_{n}\left[ C^{(3)}_{lequ}\right]^\ast_{33n2}\frac{V_{nb}}{V_{cb}}.
\end{align}
Here we do not consider the Wilson coefficient $C_{V_R}$, because it is explicitly lepton-flavour universal in the SMEFT formalism, up to contributions of  $\mathcal{O}(\mu_\text{EW}^4/\Lambda^4)$~\cite{Aebischer:2015fzz,Jenkins:2017jig,Alonso:2015sja,Cata:2015lta,Cirigliano:2009wk}. We shall also neglect terms proportional to the small CKM factors $V_{ub}$ and $V_{cb}$~\cite{Capdevila:2017iqn}, corresponding to $n=1$ and $n=2$, respectively. In such a case, the $b\to c\tau\nu$ transitions can only be affected by the Wilson coefficients $\left[C_{lq}^{(3)}\right]_{3323}$, $\left[C_{ledq}\right]_{3332}$, $\left[C^{(1)}_{lequ}\right]_{3332}$, and $\left[C^{(3)}_{lequ}\right]_{3332}$.

The last step is to evolve the WET Lagrangian $\mathcal{L}_\mathrm{WET}$ from $\mu_\text{EW}$ down to $\mu_b$ under the gauge group $SU(3)_C\otimes U(1)_\text{em}$, with the corresponding RGEs given schematically by
\begin{equation}\label{eq:RGElow}
\mu\frac{\dd{\overrightarrow {\cal C}}}{\dd \mu}=\left[\frac{\alpha_e}{4\pi}\gamma_\text{em}+\sum_{k=1}^{3}\left(\frac{\alpha_s}{4\pi}\right)^k\gamma_s^{(k)}\right]\cdot{\overrightarrow {\cal C}},
\end{equation}
where ${\overrightarrow{\cal C}}=(C_{V_L},\,C_{S_R},\,C_{S_L},\,C_T)$, and $\alpha_e=e^2/(4\pi)$ and $\alpha_s=g_s^2/(4\pi)$ are the electromagnetic and strong coupling constants, respectively. The non-zero elements of the one-loop electromagnetic anomalous dimension matrix $\gamma_\text{em}$ read~\cite{Sirlin:1981ie,Marciano:1993sh,Voloshin:1992sn,Gonzalez-Alonso:2017iyc,Aebischer:2017gaw,Jenkins:2017dyc}
\begin{align}
\left[\gamma_\text{em}\right]_{11} & =-4,
&
\left[\gamma_\text{em}\right]_{22} & =\frac{4}{3},
&
\left[\gamma_\text{em}\right]_{33} & =\frac{4}{3}, \nonumber
\\[2mm]
\left[\gamma_\text{em}\right]_{34} & =8,
&
\left[\gamma_\text{em}\right]_{43} & =\frac{1}{6},
&
\left[\gamma_\text{em}\right]_{44} & =-\frac{40}{9}.
\end{align}
The QCD anomalous dimension matrices $\gamma_s^{(k)}$ are known to three loops, with all the non-zero entries given by~\cite{Eichten:1989zv,Gracey:2000am,Gonzalez-Alonso:2017iyc,Aebischer:2017gaw,Jenkins:2017dyc}
\begin{align}
\left[\gamma_s^{(1)}\right]_{22}&=\left[\gamma_s^{(1)}\right]_{33}=-8,\quad \left[\gamma_s^{(1)}\right]_{44}=\frac{8}{3},\nonumber\\[2mm]
\left[\gamma_s^{(2)}\right]_{22}&=\left[\gamma_s^{(2)}\right]_{33}=\frac{4}{9}(-303+10n_f),\quad \left[\gamma_s^{(2)}\right]_{44}=\frac{4}{27}(543-26n_f),\nonumber\\[2mm]
\left[\gamma_s^{(3)}\right]_{22}&=\left[\gamma_s^{(3)}\right]_{33}=\frac{2}{81}\left[-101169+24(277+180\zeta_3)n_f+140n_f^2\right],\nonumber\\[2mm]
\left[\gamma_s^{(3)}\right]_{44}&=\frac{2}{81}\left[52555-2784\zeta_3-40(131+36\zeta_3)n_f-36n_f^2\right].\label{eq:ADlow}
\end{align}
As the reference energy scale in $b\to c\tau\nu$ transitions is at around $\mu_b\simeq5$~GeV, the RGE from $\mu_\text{EW}$ down to $\mu_b$ does not involve crossing any threshold, and the effective number of quark flavours $n_f$ can be fixed at $n_f=5$.

There exist several ready-made packages, such as {\tt Wilson}~\cite{Aebischer:2018bkb} and {\tt DsixTools}~\cite{Celis:2017hod}, to implement the evolution using the full one-loop anomalous dimension matrices as well as the tree-level matching. In our numerical analysis, we shall work at three-loop in QCD and one-loop in EW/QED, together with the same order for the corresponding coupling constants $\alpha_s$, $g$, $g'$ and $\alpha_e$.

\section{Observables in $b\to c\tau\nu$ transitions}
\label{sec:observables}

\subsection{$B\to D^{(\ast)}\tau\nu$}

There have been a lot of calculations for the differential decay rates of $B\to D^{(\ast)}\tau\nu$ in the presence of all the operators given in Eq.~\eqref{eq:WET_O}. In this paper, we shall follow the analytical expressions given in Refs.~\cite{Celis:2012dk,Datta:2012qk,Sakaki:2013bfa}, and consider the following observables:
\begin{itemize}
\item $q^2$-dependent and $q^2$-integrated ratios
  \begin{equation}\label{eq:RDRDs}
  R_{D^{(\ast)}}(q^2)=\frac{\dd\Gamma(B\to D^{(\ast)}\tau\nu)/\dd q^2}{\dd\Gamma(B\to D^{(\ast)}\ell\nu)/\dd q^2},
  \quad{\rm and}\quad
  R(D^{(\ast)})=\frac{{\cal B}(B\to D^{(\ast)}\tau\nu)}{{\cal B}(B\to D^{(\ast)}\ell\nu)},
  \end{equation}
  where, on the theoretical side, we define
  \begin{align}
  \dd\Gamma(B\to D^{(\ast)}\ell\nu)/\dd q^2&=\frac{1}{2}\left[\dd\Gamma(B\to D^{(\ast)}\mu\nu)/\dd q^2+\dd\Gamma(B\to D^{(\ast)}e\nu)/\dd q^2\right],\nonumber\\[0.2cm]
  {\cal B}(B\to D^{(\ast)}\ell\nu)&=\frac{1}{2}\left[{\cal B}(B\to D^{(\ast)}\mu\nu)+{\cal B}(B\to D^{(\ast)}e\nu)\right].\nonumber
  \end{align}
\item $\tau$ forward-backward asymmetry
  \begin{equation}\label{eq:AFB}
  A_{\rm FB}^{D^{(\ast)}}(q^2)=\frac{\left(\int_0^1-\int_{-1}^0\right)\dd\cos\theta\left[\dd^2\Gamma(B\to D^{(\ast)}\tau\nu)/\left(\dd q^2\dd\cos\theta\right)\right]}{\dd\Gamma(B\to D^{(\ast)}\tau\nu)/\dd q^2},
  \end{equation}
  where $\theta$ is the angle between the three-momenta of the $\tau$ lepton and the $B$ meson in the $\tau\nu$ rest frame.
\item $\tau$ spin polarization
  \begin{equation}\label{eq:spinp}
  P_\tau^{D^{(\ast)}}(q^2)=\frac{\dd\Gamma^{\lambda_\tau=1/2}(B\to D^{(\ast)}\tau\nu)/\dd q^2-\dd\Gamma^{\lambda_\tau=-1/2}(B\to D^{(\ast)}\tau\nu)/\dd q^2}{\dd\Gamma(B\to D^{(\ast)}\tau\nu)/\dd q^2},
  \end{equation}
  which can be inferred from the distinctive $\tau$ decay patterns.
\item $D^\ast$ longitudinal and transverse polarizations
  \begin{equation}\label{eq:LDsp}
  P_{\rm L}^{D^\ast}(q^2)=\frac{\dd\Gamma^{\lambda_{D^\ast}=0}(B\to D^\ast\tau\nu)/\dd q^2}{\dd\Gamma(B\to D^\ast\tau\nu)/\dd q^2},
  \quad{\rm and}\quad
  P_{\rm T}^{D^\ast}(q^2)=1-P_{\rm L}^{D^\ast}(q^2),
  \end{equation}
  which can be measured by fitting to the double differential decay distribution or from the $D^{\ast}$ decays.
\end{itemize}
Integrating separately the numerator and denominator in Eqs.~\eqref{eq:AFB}--\eqref{eq:LDsp} over the whole interval of the momentum transfer squared, $m_\tau^2\leq q^2\leq (m_B-m_{D^{(\ast)}})^2$, we can get the $q^2$-integrated observables $A_{\rm FB}(D^{(\ast)})$, $P_\tau(D^{(\ast)})$, $P_{\rm L}(D^\ast)$, and $P_{\rm T}(D^\ast)$, respectively.

In analogy to the ratios $R(D^{(\ast)})$, we can also define the following observables with the denominators involving only the light-lepton modes:
\begin{itemize}
    \item $\tau$ forward and backward fractions
    \begin{equation}
    {\cal X}_{1,2}(D^{(\ast)})=\frac{1}{2}R(D^{(\ast)})\left[1\pm A_{\rm FB}(D^{(\ast)})\right].
    \end{equation}
    \item $\tau$ spin $1/2$ and $-1/2$ fractions
    \begin{equation}
    {\cal X}_{3,4}(D^{(\ast)})=\frac{1}{2}R(D^{(\ast)})\left[1\pm P_\tau(D^{(\ast)})\right].
    \end{equation}
    \item $D^\ast$ longitudinal and transverse polarization fractions
    \begin{align}
    &{\cal X}_5(D^\ast)=R(D^\ast)P_{\rm L}(D^\ast),\\[2mm]
    &{\cal X}_6(D^\ast)=R(D^\ast)P_{\rm T}(D^\ast)=R(D^\ast)\left[1-P_{\rm L}(D^\ast)\right].
    \end{align}
\end{itemize}
It is important to note that in our scenario (\textit{i.e.} only the third-generation leptons are affected by the NP contributions) these observables are not independent. However, because of the different normalization and systematics, future measurements of them would provide important information on the size and nature of NP in $B\to D^{(\ast)}\tau\nu$ decays.

In our calculation, the $B\to D^{(\ast)}$ transition form factors are taken from Ref.~\cite{Bernlochner:2017jka}, in which both ${\cal O}(\Lambda_\text{QCD}/m_{b,c})$ and ${\cal O}(\alpha_s)$ corrections in the heavy quark effective theory are included.

\subsection{$\Lambda_b\to \Lambda_c\tau\nu$}

For an unpolarized $\Lambda_b$, the two-fold angular distribution for $\Lambda_b\to \Lambda_c\tau\nu$ can be written as~\cite{Gutsche:2015mxa,Li:2016pdv,Datta:2017aue}
\begin{equation}\label{eq:Lamb2c}
\frac{\dd^2\Gamma}{\dd q^2\dd\cos\theta_\tau}=\frac{G_F^2|V_{cb}|^2}{2}\frac{v^2|\mathbf{p}_{\Lambda_c}|}{256\pi^3m_{\Lambda_b}^2}\sum_{\lambda_{\Lambda_c}}\sum_{\lambda_\tau}\frac{1}{2}\sum_{\lambda_{\Lambda_b}}|{\cal M}_{\lambda_{\Lambda_b}}^{\lambda_{\Lambda_c},\lambda_\tau}|^2,
\end{equation}
where $v=\sqrt{1-m_\tau^2/q^2}$, and $|\mathbf{p}_{\Lambda_c}|=\sqrt{Q_+Q_-}/(2m_{\Lambda_b})$ is the magnitude of the $\Lambda_c$ three-momentum in the $\Lambda_b$ rest frame, with $Q_\pm=(m_{\Lambda_b}\pm m_{\Lambda_c})^2-q^2$, while $\theta_\tau$ is the angle between the three-momenta of the $\tau$ lepton and the $\Lambda_c$ baryon in the $\tau\nu$ rest frame. The helicity amplitudes $\mathcal{M}_{\lambda_{\Lambda_b}}^{\lambda_{\Lambda_c},\lambda_\tau}$, with the indices $\lambda_{\Lambda_b}$, $\lambda_{\Lambda_c}$ and $\lambda_\tau$ denoting respectively the helicities of the $\Lambda_b$, $\Lambda_c$ baryons and the $\tau$ lepton, can be calculated by following the helicity method described in Refs.~\cite{Korner:1987kd,Korner:1989ve,Korner:1989qb,Gratrex:2015hna}; for convenience, their explicit expressions are given in the appendix.

The observables of this process we are considering include
\begin{itemize}
  \item $q^2$-dependent and $q^2$-integrated ratios
  \begin{equation}\label{eq:RLambdac}
    R_{\Lambda_c}(q^2)=\frac{\dd\Gamma(\Lambda_b\to \Lambda_c\tau\nu)/\dd q^2}{\dd\Gamma(\Lambda_b\to \Lambda_c\mu\nu)/\dd q^2},
    \quad \text{and} \quad
    R(\Lambda_c)=\frac{{\cal B}(\Lambda_b\to \Lambda_c\tau\nu)}{{\cal B}(\Lambda_b\to \Lambda_c\mu\nu)}.
  \end{equation}
  \item $\tau$ forward-backward asymmetry
  \begin{equation}\label{eq:LamcAFB}
  A_{\rm FB}^{\Lambda_c}(q^2)=\frac{\left(\int_{0}^{1}-\int_{-1}^{0}\right)\dd\cos\theta_\tau \left[\dd^2\Gamma(\Lambda_b\to \Lambda_c\tau\nu)/(\dd q^2\dd\cos\theta_\tau)\right]}{\dd\Gamma(\Lambda_b\to \Lambda_c\tau\nu)/\dd q^2}.
  \end{equation}
  \item $\tau$ spin polarization
  \begin{equation}
    P_\tau^{\Lambda_c}(q^2)=\frac{\dd\Gamma^{\lambda_\tau=1/2}(\Lambda_b\to \Lambda_c\tau\nu)/\dd q^2-\dd\Gamma^{\lambda_\tau=-1/2}(\Lambda_b\to \Lambda_c\tau\nu)/\dd q^2}{\dd\Gamma(\Lambda_b\to \Lambda_c\tau\nu)/\dd q^2}.
  \end{equation}
  \item $\Lambda_c$ spin polarization
  \begin{equation}\label{eq:PLamc}
    P_{\Lambda_c}(q^2)=\frac{\dd\Gamma^{\lambda_{\Lambda_c}=1/2}(\Lambda_b\to \Lambda_c\tau\nu)/\dd q^2-\dd\Gamma^{\lambda_{\Lambda_c}=-1/2}(\Lambda_b\to \Lambda_c\tau\nu)/\dd q^2}{\dd\Gamma(\Lambda_b\to \Lambda_c\tau\nu)/\dd q^2}.
  \end{equation}
\end{itemize}
Integrating separately the numerator and denominator in Eqs.~\eqref{eq:LamcAFB}--\eqref{eq:PLamc} over the whole interval $q^2\in[m_\tau^2,\,(m_{\Lambda_b}-m_{\Lambda_c})^2]$, we can get the $q^2$-integrated observables $A_{\rm FB}(\Lambda_c)$, $P_\tau(\Lambda_c)$ and $P_{\Lambda_c}$, respectively. As in the mesonic case, we can also construct the following observables normalized by the corresponding muonic mode:
\begin{itemize}
\item $\tau$ forward and backward fractions
    \begin{equation}
    {\cal X}_{1,2}(\Lambda_c)=\frac{1}{2}R(\Lambda_c)\left[1\pm A_{\rm FB}(\Lambda_c)\right].
    \end{equation}
    \item $\tau$ spin $1/2$ and $-1/2$ fractions
    \begin{equation}
    {\cal X}_{3,4}(\Lambda_c)=\frac{1}{2}R(\Lambda_c)\left[1\pm P_\tau(\Lambda_c)\right].
    \end{equation}
    \item $\Lambda_c$ spin $1/2$ and $-1/2$ fractions
    \begin{equation}
    {\cal X}_{5,6}(\Lambda_c)=\frac{1}{2}R(\Lambda_c)\left[1\pm P_{\Lambda_c}\right].
    \end{equation}
  \end{itemize}

In our numerical analysis, we use the $\Lambda_b\to\Lambda_c$ transition form factors computed in lattice QCD including all the types of NP currents~\cite{Datta:2017aue,Detmold:2015aaa}.

\subsection{The rest observables}

In this subsection, we introduce the rest observables relevant for $B_c\to (J/\psi,\,\eta_c)\tau\nu$, $B\to X_c\tau\nu$ and $B_c\to \tau\nu$ decays, which could provide additional constraints on the NP parameters.

\subsubsection{$B_c\to (J/\psi,\,\eta_c)\tau\nu$}

Similar to the definitions of $R(D^{(\ast)})$, the ratios $R(J/\psi)$ and $R(\eta_c)$ for $B_c\to (J/\psi,\,\eta_c)\tau\nu$ decays are defined, respectively, by
\begin{equation}\label{eq:RJpsi}
  R(J/\psi)=\frac{{\cal B}(B_c\to J/\psi\tau\nu)}{{\cal B}(B_c\to J/\psi \mu\nu)},\qquad
  R(\eta_c)=\frac{{\cal B}(B_c\to \eta_c\tau\nu)}{{\cal B}(B_c\to \eta_c \mu\nu)}.
\end{equation}
Using the model-dependent calculations of $B_c\to (J/\psi,\,\eta_c)$ transition form factors~\cite{Ivanov:2000aj,Ebert:2003cn,Hernandez:2006gt,Ivanov:2006ni,Wang:2008xt,Qiao:2012vt,Wen-Fei:2013uea,Rui:2016opu,Issadykov:2018myx,Kiselev:1999sc,Dutta:2017xmj,Watanabe:2017mip,Tran:2018kuv}, the SM central values of $R(J/\psi)$ and $R(\eta_c)$ vary within the ranges $0.24-0.30$ and $0.25-0.35$, respectively, with the former being lower than the LHCb measurement $R(J/\psi)^\text{exp}=0.71(17)(18)$~\cite{Aaij:2017tyk} by $1.7\sigma$. Recently, model-independent bounds on $R(J/\psi)$~\cite{Cohen:2018dgz,Murphy:2018sqg,Wang:2018duy} and $R(\eta_c)$~\cite{Murphy:2018sqg,Berns:2018vpl,Wang:2018duy} are also obtained by constraining the transition form factors through a combination of dispersive relations, heavy-quark relations at zero-recoil, and the limited existing form-factor determinations from lattice QCD~\cite{Colquhoun:2016osw,Lytle:2016ixw}, resulting in $0.20\leq R(J/\psi)\leq 0.39$~\cite{Cohen:2018dgz} and $0.24\leq R(\eta_c)\leq 0.34$~\cite{Berns:2018vpl}, both of which agree with the weighted averages of previous model predictions. Here we shall use the $B_c\to (J/\psi,\,\eta_c)$ transition form factors calculated in Ref.~\cite{Wang:2008xt}, which are consistent with the preliminary lattice QCD results~\cite{Colquhoun:2016osw,Lytle:2016ixw} at all available $q^2$ points, but would result in lower central values of $R(J/\psi)$ and $R(\eta_c)$~\cite{Huang:2018nnq}.

\subsubsection{$B\to X_c\tau\nu$}

For the inclusive decay $B\to X_c\tau\nu$, we consider the ratio
\begin{equation}\label{eq:RXc}
  R(X_c)=\frac{{\cal B}(B\to X_c\tau\nu)}{{\cal B}(B\to X_c\ell\nu)}.
\end{equation}
The analytic expression of the total decay width within the SM is given by~\cite{Mannel:2017jfk}
\begin{align}\label{eq:B2XcSM}
\Gamma^\text{SM}(B\to X_c\tau\nu)= &\Gamma_0\,S_\text{em}^2\,\bigg[C_0^{(0)}+\frac{\alpha_s}{\pi}C_0^{(1)}+C_{\mu_\pi^2}\frac{\mu_\pi^2}{m_b^2}
+C_{\mu_G^2}\frac{\mu_G^2}{m_b^2}+C_{\rho_D^3}\frac{\rho_D^3}{m_b^3}+C_{\rho_{LS}^3}\frac{\rho_{LS}^3}{m_b^3}\bigg],
\end{align}
where $\Gamma_0=\frac{G_F^2|V_{cb}|^2m_b^5}{192\pi^2}$, and $S_\text{em}$ accounts for the short-distance electromagnetic correction to the SM four-fermion operator mediating the semi-leptonic decay~\cite{Sirlin:1981ie,Marciano:1993sh}. The coefficients $C_0^{(0)}$ and $C_0^{(1)}$ represent the partonic-level contributions with the leading- and  next-to-leading-order corrections in $\alpha_s$, respectively; while $C_{\mu_\pi^2}$, $C_{\mu_G^2}$ and $C_{\rho_D^3}$, $C_{\rho_{LS}^3}$ account for contributions from the $1/m_b^2$ and $1/m_b^3$ corrections in the heavy-quark expansion, respectively. Explicit analytic expressions of $C_0^{(0)}$, $C_{\mu_\pi^2}$, $C_{\mu_G^2}$ and $C_{\rho_D^3}$ can be found, for example, in Refs.~\cite{Falk:1994gw,Balk:1993sz,Mannel:2017jfk}, whereas $C_{\rho_{LS}^3}\equiv0$~\cite{Mannel:2017jfk}. The result of $C_0^{(1)}$ can be deduced, on the other hand, from Refs.~\cite{Czarnecki:1994bn,Jezabek:1996db,Biswas:2009rb}. The non-perturbative parameters $\mu_\pi^2$, $\mu_G^2$ and $\rho_D^3$, $\rho_{LS}^3$ are defined in terms of the forward matrix elements of dimension-five and -six operators, respectively. To calculate the ratio $R(X_c)$, we take~\cite{Alberti:2014yda,Bhattacharya:2018kig}: $\mu_\pi^2=0.464(67)~\text{GeV}^2$, $\mu_G^2=0.333(61)~\text{GeV}^2$, $\rho_D^3=0.175(40)~\text{GeV}^3$, and $\rho_{LS}^3=-0.146(96)~\text{GeV}^3$, with $m_b^\text{kin}(1\text{GeV})=4.561(21)~\text{GeV}$ and $m_c^\text{kin}(1\text{GeV})=1.092(20)~\text{GeV}$ in the kinetic scheme~\cite{Bigi:1996si}; the correlations between these parameters~\cite{Alberti:2014yda,Bhattacharya:2018kig} are also considered.

To discuss the NP effects from Eq.~\eqref{eq:LangWET} on the inclusive $B\to X_c\tau\nu$ decay, we take the partonic-level approximation, and decompose the decay width as~\cite{Goldberger:1999yh}
\begin{equation}
\Gamma(B\to X_c\tau\nu)=\Gamma^\text{SM}+\Gamma^\text{NP}_{(1)}+\Gamma^\text{NP}_{(2)},
\end{equation}
where the first term arises solely from the SM and is given by Eq.~\eqref{eq:B2XcSM}, while $\Gamma^\text{NP}_{(1)}$ and $\Gamma^\text{NP}_{(2)}$ represent respectively the interference term with the SM as well as the term that is of second order in the NP couplings, explicit expressions of which are taken from Ref.~\cite{Goldberger:1999yh}. Some recent works, discussing NP effects in this inclusive mode, can be found in Refs.~\cite{Grossman:1994ax,Freytsis:2015qca,Colangelo:2016ymy,Li:2016vvp,Celis:2016azn,Mannel:2017jfk,Kamali:2018fhr}.

\subsubsection{$B_c\to \tau\nu$}

The decay $B_c\to\tau\nu$, despite being at the moment out of the experimental reach~\cite{Gouz:2002kk}, can provide a powerful constraint on NP scenarios involving scalar operators~\cite{Li:2016vvp,Alonso:2016oyd,Celis:2016azn,Akeroyd:2017mhr}. In terms of the WET Lagrangian given by Eq.~\eqref{eq:LangWET}, its branching ratio can be written as
\begin{align}
{\cal B}(B_c\to\tau\nu)\;=\; & \tau_{B_c}\,\frac{m_{B_c}m_\tau^2f_{B_c}^2G_F^2|V_{cb}|^2}{8\pi}\left(1-\frac{m_\tau^2}{m_{B_c}^2}\right)^2\nonumber
\\
&\times \left|1+C_{V_L}-C_{V_R}+\frac{m_{B_c}^2}{m_\tau(\overline{m}_b+\overline{m}_c)}(C_{S_R}-C_{S_L})\right|^2,
\end{align}
where $\overline{m}_b$ and $\overline{m}_c$ are the bottom- and charm-quark running masses in the $\overline{\text{MS}}$ scheme evaluated at the scale $\mu_b$. In our numerical analysis, we take as input the lifetime $\tau_{B_c}=0.507(9)~\text{ps}$, the mass $m_{B_c}=6.2751(10)~\text{GeV}$, and the decay constant $f_{B_c}=0.434(15)~\text{GeV}$~\cite{Colquhoun:2015oha}.

An upper bound obtained from the LEP data, ${\cal B}(B_c\to\tau\nu)\lesssim 10\%$~\cite{Akeroyd:2017mhr}, is stronger than the conservative constraint, ${\cal B}(B_c\to\tau\nu)\lesssim 30\%$~\cite{Alonso:2016oyd}, by demanding that the rate does not exceed the fraction of the total width allowed by the calculation of the $B_c$ lifetime within the SM. Here we shall use the former in our numerical analysis.

\section{Numerical results and discussions}
\label{sec:results}

Before presenting the numerical results, we firstly collect in Table~\ref{tab:input} the remaining theoretical input parameters used throughout this paper. The CKM parameters are taken from Ref.~\cite{Koppenburg:2017mad}, in which the leptonic and semi-leptonic decays involving the $\mu$ and $\tau$ leptons have been removed from the global fit to the CKM parameters, following the current experimental indications that the electronic modes are in agreement with the SM predictions.

%%%%%%%%%%%%%%%%%%%%%%%%%%%%%%%%%
\begin{table}[t]
\tabcolsep 0.25in
\renewcommand\arraystretch{1.2}
\begin{center}
\caption{\label{tab:input} \small Summary of the remaining theoretical input parameters used throughout this paper.}
\vspace{0.18cm}
{\footnotesize
\begin{tabular}{|c|c|c|c|c|}
\hline\hline
\multicolumn{5}{|l|}{QCD and electroweak parameters~\cite{Tanabashi:2018oca}}\\
\hline
$G_F[10^{-5}~\text{GeV}^{-2}]$ & $\alpha_s(M_Z)$ & $\alpha_e(M_W)$ & $M_Z[\text{GeV}]$ & $\sin^2\theta_W$\\
\hline
$1.1663787(6)$ & $0.1181(11)$ & $1/128$ & $91.1876(21)$ & $0.23122(4)$\\
\hline
\multicolumn{5}{|l|}{Quark and lepton masses [GeV]~\cite{Tanabashi:2018oca}}\\
\hline
$\overline{m}_b(\overline{m}_b)$ & $\overline{m}_c(\overline{m}_c)$ & $m_\tau$ & $m_\mu$ & $m_e$\\
\hline
$4.18^{+0.04}_{-0.03}$ & $1.275^{+0.025}_{-0.035}$ & $1.77686(12)$ & $0.10566$ & $5.10999\times10^{-4}$\\
\hline
\multicolumn{5}{|l|}{CKM parameters~\cite{Koppenburg:2017mad}}\\
\hline
$\lambda$ & $A$ & $\overline{\rho}$ & \multicolumn{2}{c|}{$\overline{\eta}$}\\
\hline
$0.2251(4)$ & $0.831^{+0.021}_{-0.031}$ & $0.155(8)$ & \multicolumn{2}{c|}{$0.340(10)$}\\
\hline \hline
\end{tabular}
}
\end{center}
\end{table}
%%%%%%%%%%%%%%%%%%%%%%%%%%%%%%%%%

\subsection{Numerical effects of evolution and matching}

In this subsection, we illustrate the numerical effects of the evolution and matching procedure, based on Eqs.~\eqref{eq:RGEhigh}--\eqref{eq:betahigh} and \eqref{eq:RGElow}--\eqref{eq:ADlow}. To this end, we firstly calculate the couplings $\alpha_s$, $g$ and $g'$ at the initial scale $\Lambda$ via their RGEs within the SM. Using Eqs.~\eqref{eq:RGEhigh}--\eqref{eq:betahigh}, we can then obtain the values of $\left[ C_{lq}^{(3)}\right]_{3323}$, $\left[ C_{ledq}\right]_{3332}$, $\left[C^{(1)}_{lequ}\right]_{3332}$ and $\left[ C^{(3)}_{lequ}\right]_{3332}$ at the scale $\mu_\text{EW}$. Performing the tree-level matching at the scale $\mu_\text{EW}$, we can obtain the values of the Wilson coefficients associated with the WET operators, which can be finally run down to the scale $\mu_b$ by using Eqs.~\eqref{eq:RGElow}--\eqref{eq:ADlow}. Numerically, we have the following relations (for simplicity, the Wilson coefficients are all assumed to be real):
\begin{align}
C_{V_L}(\mu_b) & =-1.503\left[ C_{lq}^{(3)}\right]_{3323}(\Lambda),\label{eq:CVLmub}\\[2mm]
C_{V_R}(\mu_b) & =0,\\[2mm]
C_{S_L}(\mu_b) & =-1.257\left[ C^{(1)}_{lequ}\right]_{3332}(\Lambda)+0.2076\left[ C^{(3)}_{lequ}\right]_{3332}(\Lambda),\label{eq:CSLmub}\\[2mm]
C_{S_R}(\mu_b) & =-1.254\left[ C_{ledq}\right]_{3332}(\Lambda),\label{eq:CSRmub}\\[2mm]
C_{T}(\mu_b) & =0.002725\left[ C^{(1)}_{lequ}\right]_{3332}(\Lambda)-0.6059\left[ C^{(3)}_{lequ}\right]_{3332}(\Lambda),\label{eq:CTmub}
\end{align}
with $\mu_b=4.18~\text{GeV}$ and $\Lambda=1~\text{TeV}$. At the same time, the SM effective Lagrangian $\mathcal{L}_\text{SM}^{(6)}$ given by Eq.~\eqref{eq:WETSM} should be changed to $S_\text{em}\mathcal{L}_\text{SM}^{(6)}$, with $S_\text{em}\simeq1.0075$ encoding the short-distance electromagnetic correction to the SM four-fermion operator~\cite{Sirlin:1981ie,Marciano:1993sh}. It can be clearly seen from Eqs.~\eqref{eq:CSLmub} and \eqref{eq:CTmub} that there exists a large mixing of the tensor operator into the (pseudo)scalar ones under EW/QED interactions~\cite{Feruglio:2016gvd,Feruglio:2017rjo,Gonzalez-Alonso:2017iyc}.

The numerical relations given by Eqs.~\eqref{eq:CVLmub}--\eqref{eq:CTmub} allow us to connect the values of the SMEFT Wilson coefficients at the NP scale $\Lambda=1~\text{TeV}$ to that of the WET ones at the scale $\mu_b=4.18~\text{GeV}$. In order to directly use the theoretical expressions of the observables listed in section \ref{sec:observables}, which are all given in terms of the WET Wilson coefficients at the scale $\mu_b$, we need only replace Eq.~\eqref{eq:CVLmub} by $C_{V_L}(\mu_b)=(S_\text{em}-1)-1.503\left[C_{lq}^{(3)}\right]_{3323}(\Lambda)$, making the EW/QED evolution of the SM four-fermion operator also taken into account. In the following discussions, we shall use the abbreviations 
\begin{align}\label{eq:cidef}
&C_1\equiv\left[ C_{lq}^{(3)}\right]_{3323}(\Lambda),\qquad C_2\equiv\left[ C_{ledq}\right]_{3332}(\Lambda),\nonumber\\[2mm]
&C_3\equiv\left[C^{(1)}_{lequ}\right]_{3332}(\Lambda),\qquad C_4\equiv\left[ C^{(3)}_{lequ}\right]_{3332}(\Lambda),
\end{align}
for the sake of brevity.

\subsection{SM results and comparison with data}
\label{subsec:resSM}

%%%%%%%%%%%%%%%%%%%%%%%%%%%%%%%%%
\begin{table}[t]
     \tabcolsep 0.114in
	\renewcommand\arraystretch{1.2}
	\begin{center}
		\caption{\label{tab:smres}\small Predictions for the observables listed in section \ref{sec:observables} within the SM.}
		\vspace{0.18cm}
		{\footnotesize
		\begin{tabular}{|c|c|c|c|c|c|c|}
			\hline\hline
			\multicolumn{7}{|l|}{$B\to D\tau\nu$}\\
			\hline
			$R(D)^\text{SM}$ & $A_\text{FB}(D)^\text{SM}$ & $P_\tau(D)^\text{SM}$ & ${\cal X}_1(D)^\text{SM}$ & ${\cal X}_2(D)^\text{SM}$ & ${\cal X}_3(D)^\text{SM}$ & ${\cal X}_4(D)^\text{SM}$\\
			\hline
			0.2989(33) & 0.3597(3) & 0.3222(22) & 0.2032(22) & 0.0957(11) & 0.1976(24) & 0.1013(10)\\
			\hline
			\multicolumn{7}{|l|}{$B\to D^\ast\tau\nu$}\\
			\hline
			$R(D^\ast)^\text{SM}$ & $A_\text{FB}(D^\ast)^\text{SM}$ & $P_\tau(D^\ast)^\text{SM}$ & $P_\text{L}(D^\ast)^\text{SM}$ & ${\cal X}_1(D^\ast)^\text{SM}$ & ${\cal X}_2(D^\ast)^\text{SM}$ & ${\cal X}_3(D^\ast)^\text{SM}$\\
			\hline
			0.2572(29) & $-$0.0559(22) & $-$0.5039(37) & 0.4552(31) & 0.1214(13) & 0.1358(17) & 0.0638(8)\\
			\hline
			${\cal X}_4(D^\ast)^\text{SM}$ & ${\cal X}_5(D^\ast)^\text{SM}$ & ${\cal X}_6(D^\ast)^\text{SM}$ & \multicolumn{4}{l|}{}\\
			\cline{1-3}
			0.1934(23) & 0.1171(14) & 0.1401(19) & \multicolumn{4}{l|}{}\\
			\hline
			\multicolumn{7}{|l|}{$\Lambda_b\to\Lambda_c\tau\nu$}\\
			\hline
			$R(\Lambda_c)^\text{SM}$ & $A_\text{FB}(\Lambda_c)^\text{SM}$ & $P_\tau(\Lambda_c)^\text{SM}$ & $P_{\Lambda_c}^\text{SM}$ & ${\cal X}_1(\Lambda_c)^\text{SM}$ & ${\cal X}_2(\Lambda_c)^\text{SM}$ & ${\cal X}_3(\Lambda_c)^\text{SM}$\\
			\hline
			0.3328(101) & 0.0244(76) & $-$0.3077(139) & $-$0.7588(125) & 0.1705(49) & 0.1623(55) & 0.1152(37)\\
			\hline
			${\cal X}_4(\Lambda_c)^\text{SM}$ & ${\cal X}_5(\Lambda_c)^\text{SM}$ & ${\cal X}_6(\Lambda_c)^\text{SM}$ & \multicolumn{4}{l|}{}\\
			\cline{1-3}
			0.2176(75) & 0.0401(27) & 0.2927(86) &\multicolumn{4}{l|}{}\\
			\hline
			\multicolumn{7}{|l|}{The rest observables}\\
			\hline
			\multicolumn{2}{|c|}{$R(J/\psi)^\text{SM}$} & \multicolumn{2}{c|}{$R(\eta_c)^\text{SM}$} & $R(X_c)^\text{SM}$ & \multicolumn{2}{c|}{${\cal B}(B_c\to\tau\nu)^\text{SM}$}\\
			\hline
			\multicolumn{2}{|c|}{$0.2483^{+0.0060}_{-0.0055}$} & \multicolumn{2}{c|}{$0.2813^{+0.0181}_{-0.0153}$} & 0.2138(44) & \multicolumn{2}{c|}{$2.37^{+0.21}_{-0.24}\%$}\\
			\hline \hline
		\end{tabular}
	    }
	\end{center}
\end{table}
%%%%%%%%%%%%%%%%%%%%%%%%%%%%%%%%%

Our predictions for the observables listed in section \ref{sec:observables} within the SM are collected in Table~\ref{tab:smres}. The values of observables for $B\to D^{(\ast)}\tau\nu$ decays are always obtained by averaging over the charged and neutral modes. Although the relations ${\cal X}_i({\cal H})+{\cal X}_{i+1}({\cal H})=R({\cal H})$ ($i=1,3$ for ${\cal H}$ is $D$, $D^\ast$ or $\Lambda_c$, and $i=5$ for ${\cal H}$ is $D^\ast$ or $\Lambda_c$) hold, we are still presenting all of them in Table~\ref{tab:smres}, because these observables involve different normalization and systematics and can, therefore, provide complementary information on the NP scenarios. This is clearly indicated by the reduced uncertainties of the observables ${\cal X}_i({\cal H})$ compared to that of $R({\cal H})$.

Among the observables listed in Table~\ref{tab:smres}, the following ones have been measured: $R(D)^\text{exp}=0.407(39)(24)$ and $R(D^\ast)^\text{exp}=0.306(13)(7)$ with a correlation of $-0.203$~\cite{Amhis:2018up}, $P_\tau(D^\ast)^\text{exp}=-0.38^{+0.51+0.21}_{-0.51-0.16}$~\cite{Hirose:2016wfn}, $R(J/\psi)^\text{exp}=0.71(17)(18)$~\cite{Aaij:2017tyk}, and $R(X_c)^\text{exp}=0.220(22)$\footnote{This value is obtained by using the world average for the semi-leptonic branching fractions into the light leptons, ${\cal B}(B\to X_c\ell\nu)=(10.65\pm0.16)\%$~\cite{Tanabashi:2018oca}, and an averaged constraint from LEP, ${\cal B}(b\to X\tau\nu)=(2.41\pm0.23)\%$~\cite{Tanabashi:2018oca}, which is dominated by $b\to X_c\tau\nu$ because of $|V_{ub}|^2/|V_{cb}|^2\sim 1\%$ and, after correcting for the $b\to u$ contribution that is about $2\%$ due to the larger available phase space, is reduced to ${\cal B}(b\to X_c\tau\nu)=(2.35\pm0.23)\%$~\cite{Celis:2016azn}. It should be noted that the LEP measurement corresponds to a known admixture of initial states for the weak decay~\cite{Abbaneo:2001bv}. The inclusive decay rate does, however, not depend on this admixture to leading order in $1/m_b$. The corrections to this limit are hadron-specific and only partly known~\cite{Falk:1994gw,Grossman:1994ax}.}. The differences between the experimental measurements and the SM predictions ($\Delta \text{Obs.}=\text{Obs.}^\text{exp}-\text{Obs.}^\text{SM}$) for these observables read: $\Delta R(D)=0.1081(459)$ and $\Delta R(D^\ast)=0.0488(150)$ with a correlation of $-0.199$, $\Delta P_\tau(D^\ast)=0.1239(5500)$, $\Delta R(J/\psi)=0.4617(2477)$, and $\Delta R(X_c)=0.0062(224)$. These discrepancies will be used to constrain the SMEFT Wilson coefficients.

\subsection{Constraints on the SMEFT Wilson coefficients}

%%%%%%%%%%%%%%%%%%%%%%%%%%%%%%%%%
\begin{figure}[t]
	\centering
	\includegraphics[width=0.95\textwidth,pagebox=cropbox,clip]{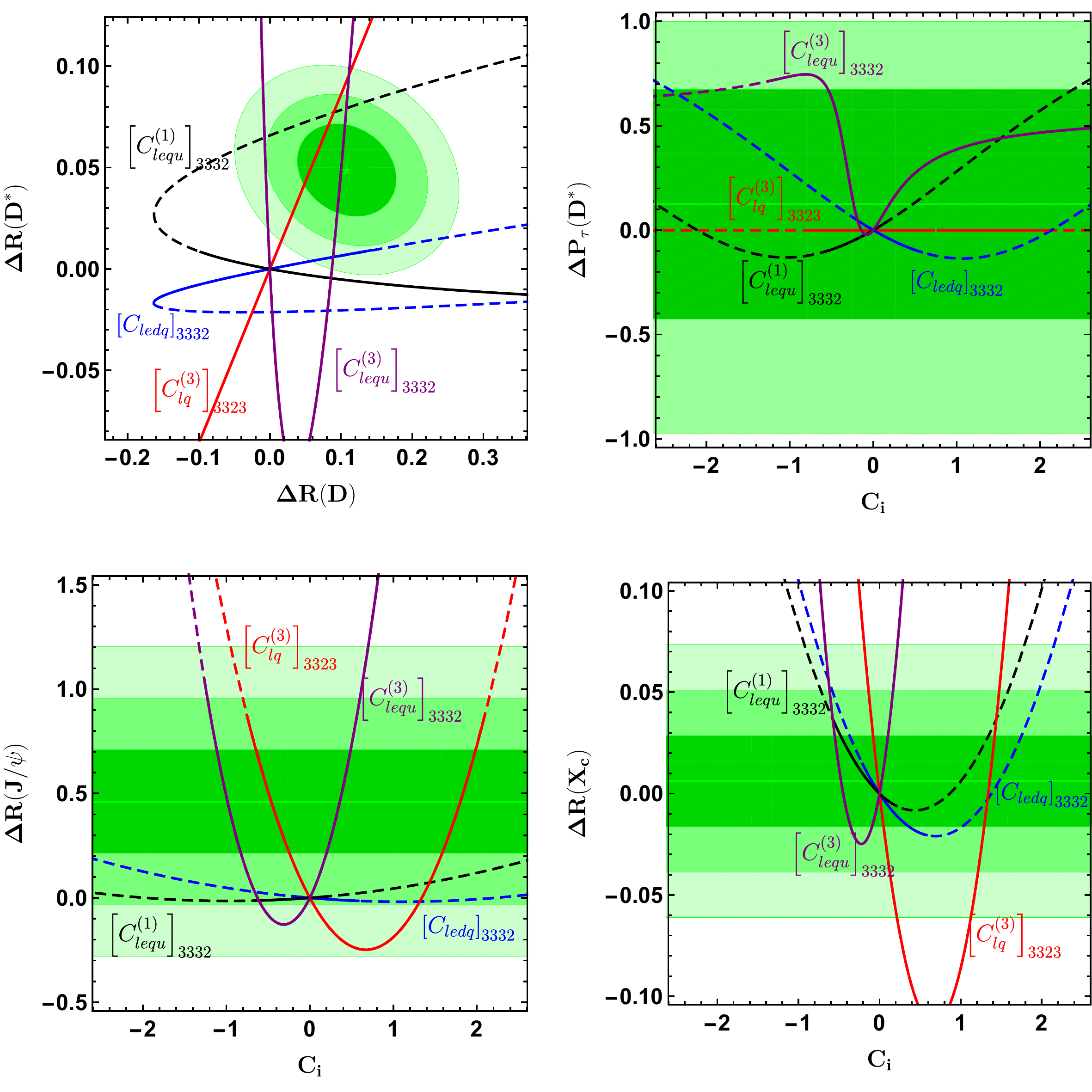}
	\caption{\label{fig:singleCi} \small Contributions to the observables in the presence of a single SMEFT Wilson coefficients $C_i$~(see Eq.~\eqref{eq:cidef}). The red, blue, black, and purple lines stand for the contributions from $C_1$, $C_2$, $C_3$, and $C_4$, respectively, with the dashed parts being already ruled out by the constraint ${\cal B}(B_c\to\tau\nu)\lesssim 10\%$. The dark-green, green, and light-green areas represent the $1$-, $2$-, and $3$-$\sigma$ differences between the measurements and the SM predictions for the observables, respectively.}
\end{figure}
%%%%%%%%%%%%%%%%%%%%%%%%%%%%%%%%%

In this subsection, we shall use $\Delta R(D)$, $\Delta R(D^\ast)$, $\Delta P_\tau(D^\ast)$, $\Delta R(J/\psi)$, and $\Delta R(X_c)$ to constrain the SMEFT Wilson coefficients $C_{1-4}$~(see Eq.~\eqref{eq:cidef}). Firstly, we show in Figure~\ref{fig:singleCi} the contributions to these observables in the presence of only a single $C_i$. It can be seen that, after taking into account the constraint ${\cal B}(B_c\to\tau\nu)\lesssim 10\%$, the scenario with a single $C_3$ is already ruled out by $\Delta R(D^{(\ast)})$ at $3\sigma$ ($99.73\%$ C.L.), and a single $C_2$ can be used to explain the $R(D^{(\ast)})$ anomalies only marginally at about $2\sigma$ ($95.45\%$ C.L.), while a single $C_1$ or $C_4$ can provide a resolution of the $R(D^{(\ast)})$ anomalies at $1\sigma$ ($68.27\%$ C.L.), especially with the finding that the central values of the current world averages of $R(D^{(\ast)})$ can be well reproduced with a single $C_4$.

Due to the large experimental uncertainty of $P_\tau(D^\ast)$, the constraint $\Delta P_\tau(D^\ast)$ on the NP Wilson coefficients is quite weak. It can be seen from the upper-right plot of Figure~\ref{fig:singleCi} that, with the constraint ${\cal B}(B_c\to\tau\nu)\lesssim 10\%$ taken into account, only $C_4$ has a significant impact on $\Delta P_\tau(D^\ast)$. Future more precise measurements of $P_\tau(D^\ast)$ at, for example, Belle II~\cite{Kou:2018nap} will be, therefore, very helpful to discriminate between the $C_1$ and $C_4$ scenarios, both of which have been found to provide reasonable explanations of the $R(D^{(\ast)})$ anomalies while satisfying the constraint ${\cal B}(B_c\to\tau\nu)\lesssim 10\%$. Using the constraints $\Delta R(J/\psi)$ and $\Delta R(X_c)$, on the other hand, we can further exclude some allowed intervals of $C_1$ and $C_4$ at $99.73\%$ C.L., which do not however affect the upper-left plot of Figure~\ref{fig:singleCi}. The constraint from $\Delta R(X_c)$ is also found to be stronger than that from $\Delta R(J/\psi)$.

\begin{figure}[t]
\centering
\includegraphics[width=0.95\textwidth,pagebox=cropbox,clip]{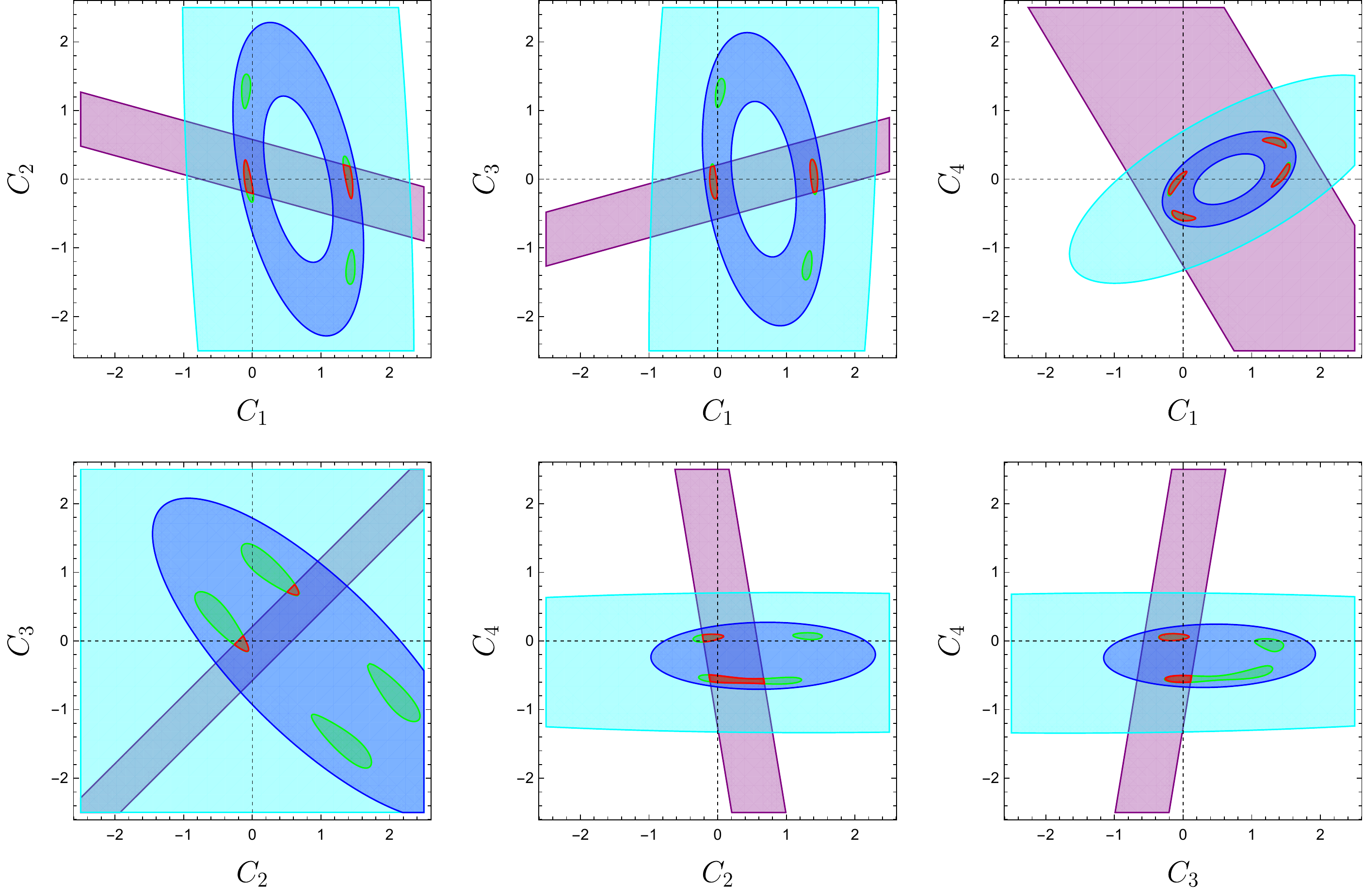}
\caption{\small Constraints on the NP Wilson coefficients in the simultaneous presence of two $C_i$s. The green, cyan, and blue areas are allowed respectively by $\Delta R(D^{(\ast)})$, $\Delta R(J/\psi)$, and $\Delta R(X_c)$, all being varied within the $3\sigma$ range of their respective experimental data, while the purple bands are allowed by the upper bound ${\cal B}(B_c\to\tau\nu)\lesssim 10\%$. The red regions are obtained by requiring $\Delta\chi^2\leq 11.83$ near the corresponding best-fit points given in Table~\ref{tab:bestfit}, under the combined constraints from the measured $R(D)$, $R(D^\ast)$, $P_\tau(D^\ast)$, $R(J/\psi)$, and $R(X_c)$, being also compatible with the ${\cal B}(B_c\to\tau\nu)\lesssim 10\%$ constraint. With such a treatment, all coloured areas (except the purple ones) correspond to $99.73\%$ C.L. regions.}
\label{fig:twoCi}
\end{figure}

In the case where two NP Wilson coefficients are present simultaneously, we show in Figure~\ref{fig:twoCi} the allowed regions in the $(C_i,C_j)$ planes under the separate constraint from $\Delta R(D^{(\ast)})$, $\Delta R(J/\psi)$, and $\Delta R(X_c)$, all being varied within $3\sigma$, as well as from the upper bound ${\cal B}(B_c\to\tau\nu)\lesssim 10\%$. As the experimental uncertainty is still quite large, we do not impose the constraint from $\Delta P_\tau(D^\ast)$ in this figure. It is found that, among all these constraints, the ones from $\Delta R(D)$ and $\Delta R(D^\ast)$ are the strongest, but the one from ${\cal B}(B_c\to\tau\nu)\lesssim 10\%$ is very complementary to them, making parts of the regions allowed by $\Delta R(D^{(\ast)})$ already excluded. It can also be seen that the ${\cal B}(B_c\to\tau\nu)$ constraint in the $(C_2,C_3)$ plane is stronger than in the other five cases.

%%%%%%%%%%%%%%%%%%%%%%%%%%%%%%%%%
%%%%%%%%%%%%%%%%%%%%%%%%%%%%%%%%%$

In order to constrain the NP Wilson coefficients $C_i$ under the combined constraints from the measured $R(D)$, $R(D^\ast)$, $P_\tau(D^\ast)$, $R(J/\psi)$, and $R(X_c)$, we construct the usual $\chi^2$ function:
\begin{align}
\chi^2(C_i)\;=\; & {\cal V}(C_i)\,\text{Cov}[\Delta R(D),\Delta R(D^\ast)]^{-1}\,{\cal V}^T(C_i)\nonumber
\\[2mm]
&+\sum_{O=P_\tau(D^\ast),\,R(J/\psi),\,R(X_c)}\frac{\left(O^\text{NP}(C_i)-\Delta O\right)^2}{\sigma^2_{\Delta O}},
\end{align}
where ${\cal V}(C_i)=[R(D)^\text{NP}(C_i)-\Delta R(D),R(D^\ast)^\text{NP}(C_i)-\Delta R(D^\ast)]$, and $\text{Cov}[\Delta R(D), \Delta R(D^\ast)]=\text{Cov}[R(D)^\text{exp}, R(D^\ast)^\text{exp}]+\text{Cov}[R(D)^\text{SM},R(D^\ast)^\text{SM}]$ is the covariance matrix between $\Delta R(D)$ and $\Delta R(D^\ast)$, the numerical value of which can be calculated by using the variance and correlation of $\Delta R(D)$ and $\Delta R(D^\ast)$ given in subsection~\ref{subsec:resSM}. Here we take in the fitting the averaged values of $R(D^{(\ast)})$
over the separate measurements by different experimental groups~\cite{Lees:2012xj,Lees:2013uzd,Huschle:2015rga,Sato:2016svk,Hirose:2016wfn,Hirose:2017dxl,Aaij:2015yra,Aaij:2017uff,Aaij:2017deq}, as compiled by HFLAV~\cite{Amhis:2018up}.%, ignoring the tensions between the separate measurements by different experimental groups~\cite{Lees:2012xj,Lees:2013uzd,Huschle:2015rga,Sato:2016svk,Hirose:2016wfn,Hirose:2017dxl,Aaij:2015yra,Aaij:2017uff,Aaij:2017deq}. Although a non-zero correlation between the individual measurements is assumed in the average, this correlation is first of all too small to be relevant in practice and secondly it is only a guesstimate by HFLAV, not officially confirmed by experiments. The more appropriate procedure to ensure the correct treatment of experimental correlations is to take the individual measurements of $R(D)$ and $R(D^{\ast})$ as separate entries, as adopted for example in Refs.~\cite{Celis:2016azn,Bhattacharya:2018kig}.

%%%%%%%%%%%%%%%%%%%%%%%%%%%%%%%%%
\begin{table}[t]
	\tabcolsep 0.142in
	\renewcommand\arraystretch{1.2}
	\begin{center}
		\caption{\small Best-fit solutions of the NP Wilson coefficients $C_i$ at the scale $\Lambda=1$~TeV, under the combined constraints from the measured $R(D)$, $R(D^\ast)$, $P_\tau(D^\ast)$, $R(J/\psi)$, and $R(X_c)$. See text for details.}
		\vspace{0.18cm}
		{\footnotesize
		\begin{tabular}{|c|c|c|c|c|}
			\hline\hline
			NP scenario & $\chi^2_{min}/$dof & best-fit point & Index & $99.73\%$ C.L. and ${\cal B}(B_c\to\tau\nu)\lesssim 10\%$ \\
			\hline
			SM $(C_i=0)$ & 23.5/5 & & &\\
			\hline
			\multirow{2}{*}{$C_1$} & $4.40/4$ & $-0.0679(150)$ & S1 & $-0.1111\to -0.0220$\\
			& $4.40/4$ & $1.408(150)$ & S2 & $1.362\to 1.452$\\
			\hline
			$C_2$ & $10.46/4$ & $-0.2198(546)$ & & $-0.2087\to -0.0423$\\
			\hline
			$C_3$ & $17.12/4$ & $-0.1814(655)$ & & $-0.3481\to 0.0425$\\
			\hline
			\multirow{2}{*}{$C_4$} & $6.51/4$ & $-0.5571(148)$ & S3 & $-0.5979\to -0.5101$\\
			& $9.14/4$ & $0.0611(149)$ & & $0.0139\to 0.1020$\\
			\hline
			\multirow{2}{*}{$(C_1,C_2)$} & $3.59/3$ & $(-0.0554, -0.0781)$ & S4 & \multirow{15}{*}{The red areas shown in Figure~\ref{fig:twoCi}}\\
			& $3.59/3$ & $(1.396, 0.0781)$ & S5 & \\
			\cline{1-4}
			\multirow{2}{*}{$(C_1,C_3)$} & $3.62/3$ & $(-0.0627, -0.0658)$ & S6 &\\
			& $3.62/3$ & $(1.403, 0.0658)$ & S7 &\\
			\cline{1-4}
			\multirow{4}{*}{$(C_1,C_4)$} & $3.79/3$ & $(-0.0962, -0.0352)$ & S8 &\\
			& $3.79/3$ & $(1.437, 0.0352)$ & S9 &\\
			& $6.51/3$ & $(0.0003, -0.5572)$ & &\\
			& $6.51/3$ & $(1.340, 0.5572)$ & &\\
			\cline{1-4}
			\multirow{2}{*}{$(C_2,C_3)$} & $3.50/3$ & $(-0.5169, 0.3580)$ & &\\
			& $3.19/3$ & $(0.1919, 1.109)$ & &\\
			\cline{1-4}
			\multirow{3}{*}{$(C_2,C_4)$} & $3.38/3$ & $(-0.1632, 0.0471)$ & S10 &\\
			& $6.06/3$ & $(0.9467, -0.5814)$ & &\\
			& $6.51/3$ & $(-0.0020, -0.5570)$ & &\\
			\cline{1-4}
			\multirow{2}{*}{$(C_3,C_4)$} & $3.38/3$ & $(-0.1643, 0.0596)$ & S11 &\\
			& $6.51/3$ & $(-0.0028, -0.5571)$ & &\\
			\hline \hline
		\end{tabular}
	    }
		\label{tab:bestfit}
	\end{center}
\end{table}
%%%%%%%%%%%%%%%%%%%%%%%%%%%%%%%%%

By minimizing the $\chi^2(C_i)$ function in different scenarios, we can get the corresponding best-fit solutions, the results of which are shown in Table~\ref{tab:bestfit}. Here the first column shows all possible cases with either a single $C_i$ or a combination of two $C_i$s, in addition to the SM case. The second column gives the values of $\chi^2_{min}$ with respect to different numbers of degrees of freedom (dof), with the corresponding best-fit points as well as the $1\sigma$ ranges ($\Delta\chi^2=\chi^2(C_i)-\chi^2_{min}\leq1$) for the single-parameter fits shown in the third column. Only the cases satisfying the condition $\chi^2_{min}\simeq \text{dof}$ are selected as the most possible solutions and are marked by different scenarios in the fourth column. In this way, we obtain eleven most trustworthy scenarios, each of which can provide a good explanation of the $R(D^{(\ast)})$ anomalies at $1\sigma$. The best-fit points allowed by $\Delta\chi^2\leq 9$ (for the single-parameter fits) or $\Delta\chi^2\leq 11.83$ (for the two-parameters fits) as well as by the upper bound ${\cal B}(B_c\to\tau\nu)\lesssim 10\%$ are finally represented in the fifth column. It is found that, after taking into account the combined constraints from $\Delta P_\tau(D^\ast)$, $\Delta R(J/\psi)$, and $\Delta R(X_c)$, the scenario with a single $C_4$ is no better than that with a single $C_1$ for resolving the $R(D^{(\ast)})$ anomalies.

\subsection{Predictions for the observables in different NP scenarios}

In order to further discriminate among the eleven most trustworthy scenarios obtained in the last subsection, we now calculate all the observables listed in section~\ref{sec:observables} within these different scenarios. Our final numerical results are collected in Tables~\ref{tab:rdrdsnpres} and \ref{tab:lamcnpres}. During the calculation, we use the central values of the NP Wilson coefficients obtained in scenarios S1 to S11, and take into account the uncertainties caused by the input parameters.

%%%%%%%%%%%%%%%%%%%%%%%%%%%%%%%%%
\begin{table}[t]
\tabcolsep 0.06in
\renewcommand\arraystretch{1.2}
\begin{center}
\caption{\small Predictions for the observables involved in $B\to D^{(\ast)}\tau\nu$ decays in all the NP scenarios.}
\vspace{0.18cm}
{\footnotesize
\begin{tabular}{|c|c|c|c|c|c|c|c|}
\hline\hline
Obs. & S1,\,S2 & S3 & S4,\,S5 & S6,\,S7 & S8,\,S9 & S10 & S11 \\
\hline
$R(D)$ & 0.3625(40) & 0.3986(42) & 0.4016(46) & 0.4007(45) & 0.3956(43) & 0.4015(48) & 0.4015(48)\\
\hline
$A_\text{FB}(D)$ & 0.3597(3) & 0.4297(4) & 0.3485(4) & 0.3504(4) & 0.3695(4) & 0.3130(5) & 0.3072(4)\\
\hline
$P_\tau(D)$ & 0.3222(22) & 0.0280(28) & 0.4087(21) & 0.3956(21) & 0.3016(23) & 0.5275(17) & 0.5364(17)\\
\hline
${\cal X}_1(D)$ & 0.2465(27) & 0.2850(30) & 0.2708(30) & 0.2706(30) & 0.2709(29) & 0.2636(31) & 0.2624(31)\\
\hline
${\cal X}_2(D)$ & 0.1161(13) & 0.1137(12) & 0.1308(16) & 0.1302(16) & 0.1247(14) & 0.1379(17) & 0.1391(18)\\
\hline
${\cal X}_3(D)$ & 0.2397(29) & 0.2049(21) & 0.2829(35) & 0.2796(34) & 0.2574(30) & 0.3067(39) & 0.3085(39)\\
\hline
${\cal X}_4(D)$ & 0.1229(12) & 0.1937(22) & 0.1187(12) & 0.1211(12) & 0.1381(14) & 0.0949(9) & 0.0931(9)\\
\hline
$R(D^\ast)$ & 0.3119(35) & 0.3042(43) & 0.3049(35) & 0.3050(35) & 0.3057(35) & 0.3057(34) & 0.3058(35)\\
\hline
$A_\text{FB}(D^\ast)$ & $-$0.0559(22) & 0.0312(15) & $-$0.0468(22) & $-$0.0634(22) & $-$0.0827(23) & 0.0010(20) & $-$0.0280(20)\\
\hline
$P_\tau(D^\ast)$ & $-$0.5039(37) & 0.1808(33) & $-$0.4867(40) & $-$0.5176(33) & $-$0.5173(42) & $-$0.4432(37) & $-$0.4973(21)\\
\hline
$P_\text{L}(D^\ast)$ & 0.4552(31) & 0.1415(13) & 0.4614(32) & 0.4501(30) & 0.4612(31) & 0.4556(32) & 0.4280(27)\\
\hline
${\cal X}_1(D^\ast)$ & 0.1473(16) & 0.1569(22) & 0.1453(16) & 0.1428(15) & 0.1402(15) & 0.1530(16) & 0.1486(16)\\
\hline
${\cal X}_2(D^\ast)$ & 0.1647(20) & 0.1474(21) & 0.1596(19) & 0.1622(20) & 0.1655(20) & 0.1527(18) & 0.1572(19)\\
\hline
${\cal X}_3(D^\ast)$ & 0.0774(9) & 0.1796(23) & 0.0783(10) & 0.0736(9) & 0.0738(10) & 0.0851(10) & 0.0769(8)\\
\hline
${\cal X}_4(D^\ast)$ & 0.2346(28) & 0.1246(21) & 0.2267(27) & 0.2315(28) & 0.2319(28) & 0.2206(27) & 0.2289(28)\\
\hline
${\cal X}_5(D^\ast)$ & 0.1420(17) & 0.0430(8) & 0.1407(18) & 0.1373(17) & 0.1410(18) & 0.1393(17) & 0.1309(15)\\
\hline
${\cal X}_6(D^\ast)$ & 0.1700(23) & 0.2612(36) & 0.1642(22) & 0.1677(22) & 0.1647(22) & 0.1664(22) & 0.1749(23)\\
\hline \hline
\end{tabular}
}
\label{tab:rdrdsnpres}
\end{center}
\end{table}
%%%%%%%%%%%%%%%%%%%%%%%%%%%%%%%%%

%%%%%%%%%%%%%%%%%%%%%%%%%%%%%%%%%
\begin{table}[t]
\tabcolsep 0.04in
\renewcommand\arraystretch{1.4}
\begin{center}
\caption{\small Predictions for the observables involved in $\Lambda_b\to\Lambda_c\tau\nu$ decay, as well as for $R(J/\psi)$, $R(\eta_c)$, $R(X_c)$, and ${\cal B}(B_c\to\tau\nu)$ in all the NP scenarios.}
\vspace{0.18cm}
{\scriptsize
\begin{tabular}{|c|c|c|c|c|c|c|c|}
\hline\hline
Obs. & S1,\,S2 & S3 & S4,\,S5 & S6,\,S7 & S8,\,S9 & S10 & S11 \\
\hline
$R(\Lambda_c)$ & 0.4037(123) & 0.3646(247) & 0.4085(125) & 0.4090(125) & 0.4107(123) & 0.4044(130) & 0.4043(133)\\
\hline
$A_\text{FB}(\Lambda_c)$ & 0.0244(76) & 0.1129(166) & 0.0393(75) & 0.0280(76) & 0.0181(81) & 0.0628(68) & 0.0412(67)\\
\hline
$P_\tau(\Lambda_c)$ & $-$0.3077(139) & 0.1002(282) & $-$0.2487(155) & $-$0.2734(155) & $-$0.3076(145) & $-$0.1791(164) & $-$0.2108(169)\\
\hline
$P_{\Lambda_c}$ & $-$0.7588(125) & 0.1170(736) & $-$0.7570(117) & $-$0.7433(117) & $-$0.8031(111) & $-$0.6833(126) & $-$0.6188(127)\\
\hline
${\cal X}_1(\Lambda_c)$ & 0.2068(59) & 0.2029(155) & 0.2123(62) & 0.2102(61) & 0.2091(59) & 0.2149(66) & 0.2104(66)\\
\hline
${\cal X}_2(\Lambda_c)$ & 0.1969(67) & 0.1617(98) & 0.1962(67) & 0.1988(68) & 0.2016(68) & 0.1895(66) & 0.1938(69)\\
\hline
${\cal X}_3(\Lambda_c)$ & 0.1397(45) & 0.2006(155) & 0.1535(53) & 0.1486(52) & 0.1422(47) & 0.1660(61) & 0.1595(61)\\
\hline
${\cal X}_4(\Lambda_c)$ & 0.2640(91) & 0.1641(114) & 0.2551(88) & 0.2604(89) & 0.2685(91) & 0.2384(85) & 0.2447(89)\\
\hline
${\cal X}_5(\Lambda_c)$ & 0.0487(32) & 0.2037(249) & 0.0496(31) & 0.0525(32) & 0.0404(28) & 0.0640(37) & 0.0770(41)\\
\hline
${\cal X}_6(\Lambda_c)$ & 0.3550(104) & 0.1610(107) & 0.3589(106) & 0.3565(105) & 0.3703(108) & 0.3404(105) & 0.3272(102)\\
\hline
$R(J/\psi)$ & $0.3012_{-0.0066}^{+0.0073}$ & $0.1980_{-0.0167}^{+0.0215}$ & $0.2939_{-0.0066}^{+0.0073}$ & $0.2949_{-0.0064}^{+0.0069}$ & $0.2935_{-0.0069}^{+0.0080}$ & $0.2953_{-0.0064}^{+0.0067}$ & $0.2971_{-0.0063}^{+0.0062}$\\
\hline
$R(\eta_c)$ & $0.3412_{-0.0185}^{+0.0219}$ & $0.3159_{-0.0261}^{+0.0304}$ & $0.3766_{-0.0227}^{+0.0268}$ & $0.3760_{-0.0223}^{+0.0264}$ & $0.3692_{-0.0181}^{+0.0216}$ & $0.3780_{-0.0277}^{+0.0324}$ & $0.3788_{-0.0285}^{+0.0332}$\\
\hline
$R(X_c)$ & 0.2381(40) & 0.2439(39) & 0.2388(39) & 0.2391(39) & 0.2405(39) & 0.2366(40) & 0.2365(40)\\
\hline
${\cal B}(B_c\to\tau\nu)$[\%] & $2.87_{-0.29}^{+0.26}$ & $5.32_{-0.54}^{+0.47}$ & $5.36_{-0.55}^{+0.48}$ & $1.29_{-0.13}^{+0.11}$ & $3.27_{-0.33}^{+0.29}$ & $8.02_{-0.82}^{+0.71}$ & $7.68_{-0.78}^{+0.68}\times10^{-3}$\\
\hline \hline
\end{tabular}
}
\label{tab:lamcnpres}
\end{center}
\end{table}
%%%%%%%%%%%%%%%%%%%%%%%%%%%%%%%%%

From Tables~\ref{tab:rdrdsnpres} and \ref{tab:lamcnpres}, we can see that the scenarios S1 and S2, S4 and S5, S6 and S7, as well as S8 and S9, all of which involve the NP Wilson coefficient $C_1$ that would induce only the left-handed vector current at the scale $\mu_b$ (see Eq.~\eqref{eq:CVLmub}), cannot be distinguished from each other. There are, however, a number of observables, such as $P_\tau(D)$, ${\cal X}_4(D)$, $A_\text{FB}(D^\ast)$, $P_\tau(D^\ast)$, $P_\text{L}(D^\ast)$, ${\cal X}_5(D^\ast)$, $P_\tau(\Lambda_c)$, $P_{\Lambda_c}$, and ${\cal X}_5(\Lambda_c)$, that can be used to distinguish the scenario S3 from the other ones. In addition to the scenario S3, there exist another two scenarios S10 and S11 that do not involve the Wilson coefficient $C_1$. As the predicted branching fraction of $B_c\to\tau\nu$ decay in the scenario S11 is much smaller than in the other scenarios as well as in the SM, we can use the observable ${\cal B}(B_c\to\tau\nu)$ to distinguish the scenario S11 from the other ones. On the other hand, the observables $P_\tau(D)$, $A_\text{FB}(D^\ast)$, $A_\text{FB}(\Lambda_c)$, $P_\tau(\Lambda_c)$, and ${\cal B}(B_c\to\tau\nu)$ have the potential to distinguish the scenario S10 from the other ones.

The scenarios S4, S5 and S6, S7 might be distinguished only by the observable ${\cal B}(B_c\to\tau\nu)$. While the observables ${\cal X}_i$ can help to distinguish the scenarios S1, S2 from the SM, the corresponding observables normalized by the tauonic modes, such as the $\tau$ forward-backward asymmetries $A_\text{FB}(D)$, $A_\text{FB}(D^\ast)$, and $A_\text{FB}(\Lambda_c)$, fail to do, because they are all identically the same in the scenarios S1 and S2 as well as in the SM.

%%%%%%%%%%%%%%%%%%%%%%%%%%%%%%%%%
\begin{figure}[htbp]
	\centering
	\includegraphics[width=0.96\textwidth,pagebox=cropbox,clip]{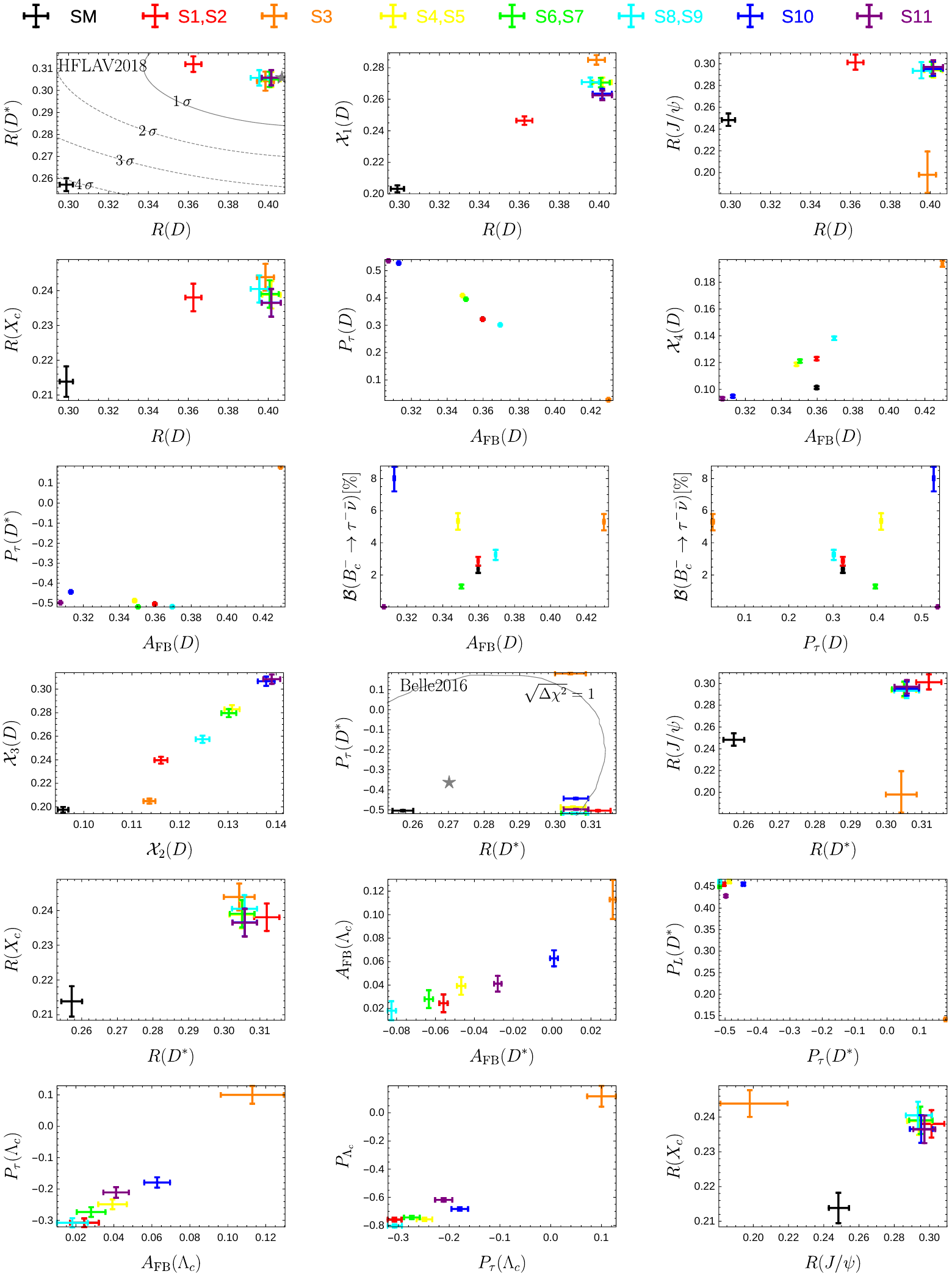}
	\caption{\small Correlatios among some of the observables discussed in this paper. Gray star point in the $R(D)-R(D^\ast)$ and $R(D^\ast)-P_\tau(D^\ast)$ plots correspond to the experimental central values.}
	\label{fig:corrplot}
\end{figure}
%%%%%%%%%%%%%%%%%%%%%%%%%%%%%%%%%

In order to further differentiate these different scenarios, we now consider the correlations among the observables discussed in this paper. There are totally 465 correlation plots, with a small part of them shown in Figure~\ref{fig:corrplot}. As can be seen from the $R(D)-R(D^\ast)$ correlation plot, it is interesting to note that all the NP scenarios can resolve the $R(D^{(\ast)})$ anomalies at $1\sigma$ very well and, except in S1 and S2, the predicted $R(D^{(\ast)})$ in the other nine scenarios are very close to the center values of the current experimental data. The $R(D)-R(D^\ast)$ and $R(D)-{\cal X}_1(D)$ correlation plots have, therefore, the potential to distinguish the scenarios S1 and S2 from the other ones. Different patterns for different NP scenarios are also observed in the other correlation plots. For example, the $A_\text{FB}(D)-P_\tau(D)$, $A_\text{FB}(D)-{\cal X}_4(D)$, ${\cal X}_2(D)-{\cal X}_3(D)$, $A_\text{FB}(D)-{\cal B}(B_c\to\tau\nu)$, and $P_\tau(D)-{\cal B}(B_c\to\tau\nu)$ correlation plots can be used to distinguish the scenarios S1 and S2 from the scenarios without $C_1$. The predicted patterns in the $P_\tau(D^\ast)-P_\text{L}(D^\ast)$ and $A_\text{FB}(D)-P_\tau(D^\ast)$ in the scenario S3 are also found to be very different from the ones in the other scenarios.

Based on all the above observations, we can, therefore, conclude that all the eleven NP scenarios, except S1 and S2, S4 and S5, S6 and S7, as well as S8 and S9, can be distinguished from each other by the above observables as well as their correlations.

\subsection{The $SU(2)_L$-invariant implications}

Due to the $SU(2)_L$ invariance of the SMEFT Lagrangian, the non-zero Wilson coefficients $C_{1-4}$ at the high-energy scale $\Lambda$ enter not only in the $b\to c\tau\nu_\tau$ processes studied in this paper but also in other low-energy charged and/or neutral current processes~\cite{Alonso:2015sja,Capdevila:2017iqn,Gonzalez-Alonso:2016etj}. With our prescription for the weak and mass eigenstates of fermion fields (see Eq.~\eqref{eq:CKMeff}), the processes $c\to d_n\tau\nu$, $d_m\to d_n \nu\bar\nu$ and $d_n\to d_m \tau^+\tau^-$, with $d_m$ being one of the $d$-type quarks in the mass eigenstate, will also receive the NP contributions from $C_{1-4}$. However, compared with the $b\to c\tau\nu$ transitions, the NP effects on $c\to d_n\tau\nu$~($d_n=d$ or $s$) processes are suppressed by the small factor $V_{cb}V_{tn}/V_{cn}\simeq 1.6\times 10^{-3}$, and we can, therefore, neglect safely the NP impacts on the $D_{(s)}$-meson decays. On the other hand, it is found that the upper bound on the branching fraction of $B^+\to K^+\nu\bar\nu$ decay given by the Belle~\cite{Lutz:2013ftz} and BaBar~\cite{Lees:2013kla} collaborations will disfavour the larger parameter regions for $C_1$ given in Table~\ref{tab:bestfit}. Combining the $\chi^2$-fit results with the constraint from the branching fraction of $B^+\to K^+\nu\bar\nu$~\cite{Altmannshofer:2009ma,Bartsch:2009qp,Buras:2014fpa,Becirevic:2015asa,Celis:2015ara,Calibbi:2015kma,Bauer:2015knc,Fajfer:2015ycq,Barbieri:2015yvd,Deshpand:2016cpw}, we also find that the $C_1$ contributions to the branching fractions of some $b\to s\tau^+\tau^-$ processes, such as $B_s\to\tau^+\tau^-$, $B\to K^{(\ast)}\tau^+\tau^-$, and $B_s\to \phi\tau^+\tau^-$ decays, can be enhanced by about two orders of magnitude compared to the SM~\cite{Capdevila:2017iqn}. The NP effects on $\Upsilon(nS)\to \tau^+\tau^-$ decays are, however, suppressed by the small factor $V_{cb}V_{ts}/V_{cs}$ compared to these $b\to s\tau^+\tau^-$ processes. Finally, it should be noted that there also exist some collider signals directly implied by the $R(D^{(\ast)})$ anomalies. For example, the partonic-level process $gc\to b\tau\nu$ implied by crossing symmetry from the $b\to c\tau\nu$ decay should also take place at the LHC~\cite{Altmannshofer:2017poe}. Furthermore, the $\tau^+\tau^-$ resonance searches at the LHC~\cite{Aad:2015osa,Aaboud:2016cre} should also be confronted with what have been found in this paper~\cite{Faroughy:2016osc}. Detailed analyses of the $SU(2)_L$-invariant implications will be presented in a forthcoming paper.

\section{Conclusions}
\label{sec:conclusion}

In this paper, we have discussed the $B\to D^{(\ast)}\tau\nu$, $\Lambda_b\to\Lambda_c\tau\nu$, $B_c\to (J/\psi,\,\eta_c)\tau\nu$, $B\to X_c\tau\nu$, and $B_c\to\tau\nu$ decays, all being mediated by the same quark-level $b\to c\tau\nu$ transition, in the SMEFT framework. First of all, we obtained the WET Lagrangian describing the $b\to c\tau\nu$ transitions at the scale $\mu_b=4.18$~GeV, in terms of the Wilson coefficients of the SMEFT operators $C_1\equiv\left[ C_{lq}^{(3)}\right]_{3323}(\Lambda)$, $C_2\equiv\left[ C_{ledq}\right]_{3332}(\Lambda)$, $C_3\equiv\left[C^{(1)}_{lequ}\right]_{3332}(\Lambda)$, and $C_4\equiv\left[ C^{(3)}_{lequ}\right]_{3332}(\Lambda)$ given at the NP scale $\Lambda=1$~TeV. This is achieved by using the RGEs at three-loop in QCD and one-loop in EW/QED based on Eqs.~\eqref{eq:RGEhigh}--\eqref{eq:betahigh} and \eqref{eq:RGElow}--\eqref{eq:ADlow}, as well as the tree-level matching of the SMEFT Lagrangian onto the WET one at the EW scale $\mu_{\rm EW}$, with the final numerical relations summarized by Eqs.~\eqref{eq:CVLmub}--\eqref{eq:CTmub}, which allow us to connect the values of the SMEFT Wilson coefficients given at $\Lambda$ to that of the WET ones given at $\mu_b$.

We then explored the contributions to the observables $R(D)$, $R(D^\ast)$, $P_\tau(D^\ast)$, $R(J/\psi)$, and $R(X_c)$ in the presence of a single SMEFT Wilson coefficient. It is found that the scenario with a single $C_1$ or $C_4$ can be used to resolve the $R(D^{(\ast)})$ anomalies at $1\sigma$, especially with the finding that the experimental central values can be well reproduced with a single $C_4$. A single $C_3$ is, however, already ruled out by the measured $R(D^{(\ast)})$ and the constraint ${\cal B}(B_c\to\tau\nu)\lesssim 10\%$ at more than $3\sigma$. In the case where two SMEFT Wilson coefficients are present simultaneously, on the other hand, we found that the constraints from $\Delta R(D)$ and $\Delta R(D^\ast)$ are the strongest, but the one from ${\cal B}(B_c\to\tau\nu)\lesssim 10\%$ is very complementary to them, making parts of the regions allowed by $\Delta R(D^{(\ast)})$ already excluded. Under the combined constraints from the measured $R(D)$, $R(D^\ast)$, $P_\tau(D^\ast)$, $R(J/\psi)$, and $R(X_c)$, we obtained the best-fit points and the allowed regions at $99.73\%$ C.L., which are shown in Table~\ref{tab:bestfit} and Figure~\ref{fig:twoCi}, respectively. Due to the extra combined constraints from $P_\tau(D^\ast)$, $R(J/\psi)$, and $R(X_c)$, the scenario with a single $C_4$ is also found to be no better than that with a single $C_1$ for resolving the $R(D^{(\ast)})$ anomalies.

Through a global fit, we have identified eleven most trustworthy scenarios, each of which can provide a good explanation of the $R(D^{(\ast)})$ anomalies at $1\sigma$. In order to further discriminate these different scenarios, we have also predicted the observables in each NP scenario and considered the correlations among them. It is found that most of the scenarios can be differentiated from each other by using these observables as well as their correlations. In particular, the predicted ${\cal B}(B_c\to\tau\nu)$ in the scenario S11 is found to be much smaller than in the other scenarios as well as in the SM. The observables $P_\tau(D)$, ${\cal X}_4(D)$, $A_\text{FB}(D^\ast)$, $P_\tau(D^\ast)$, $P_\text{L}(D^\ast)$, ${\cal X}_5(D^\ast)$, $P_\tau(\Lambda_c)$, $P_{\Lambda_c}$, ${\cal X}_5(\Lambda_c)$, as well as the $P_\tau(D^\ast)-P_\text{L}(D^\ast)$ and $A_\text{FB}(D)-P_\tau(D^\ast)$ correlation plots can be used to distinguish the scenario S3 from the other ones.

As both the LHCb and Belle II experiments will be in an ideal position to provide additional information by significantly reducing the uncertainties of the observables already measured and by measuring new observables that can provide complementary constraints on the NP parameters, we shall expect a better understanding of the different NP scenarios involved in $b\to c\tau\nu$ transitions.

\vspace{0.2cm}

{\bf Note added:} After this work was finished, we are informed that there has been a preliminary Belle measurement of the $D^\ast$ longitudinal polarization fraction in $B\to D^\ast \tau \nu$~\cite{Adamczyk:2018}. This preliminary result $P_\text{L}(D^\ast)=0.60\pm0.08\pm0.035$ shall exclude the scenario S3, which predicts a very small $P_\text{L}(D^\ast)=0.142\pm0.001$ (see Table~\ref{tab:rdrdsnpres}). This implies that the solution to the $R(D)$ and $R(D^\ast)$ anomalies with the tensor operator is not favored. 

\section*{Acknowledgements}

This work is supported by the National Natural Science Foundation of China  under Grant Nos.~11675061, 11775092, 11521064 and 11435003. X.L. is also supported in part by the self-determined research funds of CCNU from the colleges' basic research and operation of MOE~(CCNU18TS029). Q.H. is also supported by the China Postdoctoral Science Foundation~(2018M632896).

\appendix
\renewcommand{\theequation}{A.\arabic{equation}}

\section*{Appendix: Helicity amplitudes for $\Lambda_b\to \Lambda_c\tau\nu$ decay}
\label{app:HA}

Here we give the explicit expressions of the helicity amplitudes for $\Lambda_b(p_{\Lambda_b})\to \Lambda_c(p_{\Lambda_c})\tau(p_\tau)\nu(p_\nu)$ decay calculated by ourselves. Following the helicity method described in Refs.~\cite{Korner:1987kd,Korner:1989ve,Korner:1989qb,Gratrex:2015hna}, we can write the helicity amplitudes ${\cal M}_{\lambda_{\Lambda_b}}^{\lambda_{\Lambda_c},\lambda_\tau}$ as~\cite{Datta:2017aue}
\begin{equation}
{\cal M}_{\lambda_{\Lambda_b}}^{\lambda_{\Lambda_c},\lambda_\tau}=H^{(SP)\lambda_{\Lambda_c}}_{\lambda_{\Lambda_b}}L^{\lambda_\tau}+\sum_\lambda \eta_\lambda H^{(VA)\lambda_{\Lambda_c}}_{\lambda_{\Lambda_b},\lambda}L^{\lambda_\tau}_\lambda+\sum_{\lambda,\lambda'} \eta_\lambda \eta_{\lambda'} H^{(T)\lambda_{\Lambda_c}}_{\lambda_{\Lambda_b},\lambda,\lambda'}L^{\lambda_\tau}_{\lambda,\lambda'}.
\end{equation}
Here, $H$ and $L$ denote the hadronic and leptonic helicity amplitudes, respectively,  $\lambda^{(\prime)}$ indicates the helicity of the virtual vector boson, with $\eta_{\lambda^{(\prime)}}=1$ for $\lambda^{(\prime)}=t$ and $\eta_{\lambda^{(\prime)}}=-1$ for $\lambda^{(\prime)}=0,\,\pm1$, and the momentum transfer squared is given by $q^2=(p_{\Lambda_b}-p_{\Lambda_c})^2=(p_\tau+p_\nu)^2$.
						
Starting with the effective Lagrangian given by Eq.~\eqref{eq:LangWET} and using the helicity-based definition of the $\Lambda_b\to\Lambda_c$ transition form factors in Ref.~\cite{Datta:2017aue,Feldmann:2011xf}, we can obtain the hadronic helicity amplitudes as follows:
\begin{itemize}
\item The non-zero scalar and pseudo-scalar helicity amplitudes,
  \begin{align}
  H^{(SP)-1/2}_{-1/2} =
  &(C_{S_L}+C_{S_R})\,F_0\,\frac{\sqrt{Q_+}}{\overline{m}_b-\overline{m}_c}\,\mLambdam\nonumber\\
  &-(C_{S_L}-C_{S_R})\,G_0\,\frac{\sqrt{Q_-}}{\overline{m}_b+\overline{m}_c}\,\mLambdap,\\[2mm]
  H^{(SP)1/2}_{1/2} =
  &(C_{S_L}+C_{S_R})\,F_0\,\frac{\sqrt{Q_+}}{\overline{m}_b-\overline{m}_c}\,\mLambdam\nonumber\\
  &+(C_{S_L}-C_{S_R})\,G_0\,\frac{\sqrt{Q_-}}{\overline{m}_b+\overline{m}_c}\,\mLambdap,
  \end{align}
  where $\overline{m}_b$ and $\overline{m}_c$ are the $b$- and $c$-quark running masses in the $\overline{\text{MS}}$ scheme and should be evaluated at the typical energy scale $\mu_b$.

\item The non-zero vector and axial-vector helicity amplitudes,
  \begin{align}
  H^{(VA)-1/2}_{-1/2,0}=
  &(1+C_{V_L}+C_{V_R})\,F_+\,\frac{\sqrt{Q_-}}{\sqrt{q^2}}\,\mLambdap\nonumber\\
  &+(1+C_{V_L}-C_{V_R})\,G_+\,\frac{\sqrt{Q_+}}{\sqrt{q^2}}\,\mLambdam,\\[2mm]
  H^{(VA)-1/2}_{-1/2,t}=
  &(1+C_{V_L}+C_{V_R})\,F_0\,\frac{\sqrt{Q_+}}{\sqrt{q^2}}\,\mLambdam\nonumber\\
  &+(1+C_{V_L}-C_{V_R})\,G_0\,\frac{\sqrt{Q_-}}{\sqrt{q^2}}\,\mLambdap,\\[2mm]
  H^{(VA)-1/2}_{1/2,1}=&(1+C_{V_L}+C_{V_R})\,F_\perp\,\sqrt{2Q_-}+(1+C_{V_L}-C_{V_R})\,G_\perp\,\sqrt{2Q_+},\\[2mm]
  H^{(VA)1/2}_{-1/2,-1}=&(1+C_{V_L}+C_{V_R})\,F_\perp\,\sqrt{2Q_-}-(1+C_{V_L}-C_{V_R})\,G_\perp\,\sqrt{2Q_+},\\[2mm]
  H^{(VA)1/2}_{1/2,0}=
  &(1+C_{V_L}+C_{V_R})\,F_+\,\frac{\sqrt{Q_-}}{\sqrt{q^2}}\,\mLambdap\nonumber\\
  &-(1+C_{V_L}-C_{V_R})\,G_+\,\frac{\sqrt{Q_+}}{\sqrt{q^2}}\,\mLambdam,\\[2mm]
  H^{(VA)1/2}_{1/2,t}=
  &(1+C_{V_L}+C_{V_R})\,F_0\,\frac{\sqrt{Q_+}}{\sqrt{q^2}}\,\mLambdam\nonumber\\
  &-(1+C_{V_L}-C_{V_R})\,G_0\,\frac{\sqrt{Q_-}}{\sqrt{q^2}}\,\mLambdap.
  \end{align}

\item The non-zero tensor helicity amplitudes,
  \begin{align}
  H^{(T)-1/2}_{-1/2,t,0}&=H^{(T)-1/2}_{-1/2,1,-1}=C_T(h_+\sqrt{Q_-}-\widetilde{h}_+\sqrt{Q_+}),\\[2mm]
  H^{(T)-1/2}_{1/2,1,0}&=H^{(T)-1/2}_{1/2,t,1}=
  \sqrt{2}C_T\bigg[h_\perp\frac{\sqrt{Q_-}}{\sqrt{q^2}}\mLambdap -\widetilde{h}_\perp\frac{\sqrt{Q_+}}{\sqrt{q^2}}\mLambdam\bigg],\\[2mm]
  H^{(T)1/2}_{-1/2,0,-1}&=H^{(T)1/2}_{-1/2,t,-1}=
  \sqrt{2}C_T\bigg[h_\perp\frac{\sqrt{Q_-}}{\sqrt{q^2}}\mLambdap +\widetilde{h}_\perp\frac{\sqrt{Q_+}}{\sqrt{q^2}}\mLambdam\bigg],\\[2mm]
  H^{(T)1/2}_{1/2,1,-1}&=H^{(T)1/2}_{1/2,t,0}=C_T(h_+\sqrt{Q_-}+\widetilde{h}_+\sqrt{Q_+}),
  \end{align}
  together with the other non-vanishing tensor-type helicity amplitudes related to the above ones by
  \begin{equation}
  H^{(T)\lambda_{\Lambda_c}}_{\lambda_{\Lambda_b},\lambda,\lambda'}=-H^{(T)\lambda_{\Lambda_c}}_{\lambda_{\Lambda_b},\lambda',\lambda}.
  \end{equation}
\end{itemize}

For the leptonic helicity amplitudes, on the other hand, we obtain~\cite{Datta:2017aue,Tanaka:2012nw}:
\begin{itemize}
\item The non-zero scalar and pseudoscalar leptonic helicity amplitudes,
  \begin{equation}
    L^{1/2}=2\sqrt{q^2}\,v.
  \end{equation}

\item The non-zero vector and axial-vector amplitudes,
  \begin{align}
  L^{-1/2}_{0} & =2\sqrt{q^2}\,v\,\sin(\theta_\tau),\\[2mm]
  L^{-1/2}_{\pm1} & =-\sqrt{2q^2}\,v\,\left[1\mp\cos(\theta_\tau)\right],\\[2mm]
  L^{1/2}_{0} & =-2\,m_\tau \,v\,\cos(\theta_\tau),\\[2mm]
  L^{1/2}_{\pm1} & =\pm\sqrt{2}m_\tau \,v\,\sin(\theta_\tau),\\[2mm]
  L^{1/2}_{t} & =2\,m_\tau \,v.
  \end{align}
  \item The non-zero tensor amplitudes,
  \begin{align}
  L^{-1/2}_{0,\pm1} & =\mp\sqrt{2}\,m_\tau\,v\,\left[1\mp\cos(\theta_\tau)\right],\\[2mm]
  L^{-1/2}_{0,t}=L^{-1/2}_{1,-1} & =-2\,m_\tau\,v\,\sin(\theta_\tau),\\[2mm]
  L^{-1/2}_{\pm1,t} & =\sqrt{2}\,m_\tau\,v\,\left[1\mp\cos(\theta_\tau)\right],\\[2mm]
  L^{1/2}_{0,\pm1} & =\sqrt{2q^2}\,v\,\sin(\theta_\tau),\\[2mm]
  L^{1/2}_{0,t}=L^{1/2}_{1,-1} & =2\,\sqrt{q^2}\,v\,\cos(\theta_\tau),\\[2mm]
  L^{1/2}_{\pm1,t} & =\mp\sqrt{2q^2}\,v\,\sin(\theta_\tau),
  \end{align}
  as well as the other non-vanishing tensor-type helicity amplitudes related to the above ones by
  \begin{equation}
  L^{\lambda_\tau}_{\lambda,\lambda'}=-L^{\lambda_\tau}_{\lambda',\lambda}.
  \end{equation}
\end{itemize}

Integrating the two-fold angular distribution given by Eq.~\eqref{eq:Lamb2c} over $\cos\theta_\tau$ but without the first two summations over $\lambda_{\Lambda_c}$ and $\lambda_{\tau}$, we can obtain the following expression for the helicity-dependent differential decay rate:
\begin{equation}\label{eq:Lamb2cq2}
\frac{\dd\Gamma^{\lambda_{\Lambda_c},\lambda_\tau}}{\dd q^2}=\frac{G_F^2|V_{cb}|^2}{2}\frac{v^2|\mathbf{p}_{\Lambda_c}|}{256\pi^3m_{\Lambda_b}^2}\frac{1}{2}\sum_{\lambda_{\Lambda_b}}\int_{-1}^{1}\dd\cos\theta_\tau|{\cal M}_{\lambda_{\Lambda_b}}^{\lambda_{\Lambda_c},\lambda_\tau}|^2,
\end{equation}
from which we get the differential decay rates
\begin{equation}
\frac{\dd\Gamma^{\lambda_{\Lambda_c}=1/2}}{\dd q^2}=\sum_{\lambda_\tau}\frac{\dd\Gamma^{1/2,\lambda_\tau}}{\dd q^2},\quad
\frac{\dd\Gamma^{\lambda_{\Lambda_c}=-1/2}}{\dd q^2}=\sum_{\lambda_\tau}\frac{\dd\Gamma^{-1/2,\lambda_\tau}}{\dd q^2},
\end{equation}
for a polarized $\Lambda_c$ baryon, and
\begin{equation}
\frac{\dd\Gamma^{\lambda_\tau=1/2}}{\dd q^2}=\sum_{\lambda_{\Lambda_c}}\frac{\dd\Gamma^{\lambda_{\Lambda_c},1/2}}{\dd q^2},\quad
\frac{\dd\Gamma^{\lambda_\tau=-1/2}}{\dd q^2}=\sum_{\lambda_{\Lambda_c}}\frac{\dd\Gamma^{\lambda_{\Lambda_c},-1/2}}{\dd q^2},
\end{equation}
for a polarized $\tau$ lepton.

\bibliographystyle{JHEP}
\bibliography{references}

\providecommand{\href}[2]{#2}\begingroup\raggedright\begin{thebibliography}{100}

\bibitem{Bifani:2018zmi}
S.~Bifani, S.~e. Descotes-Genon, A.~Romero~Vidal, and M.-H. Schune, {\it
  {Review of Lepton Universality tests in $B$ decays}},  {\it J. Phys.} {\bf
  G46} (2019), no.~2 023001, [\href{http://arxiv.org/abs/1809.06229}{{\tt
  arXiv:1809.06229}}].

\bibitem{Li:2018lxi}
Y.~Li and C.-D. Lü, {\it {Recent Anomalies in B Physics}},  {\it Sci. Bull.}
  {\bf 63} (2018) 267--269, [\href{http://arxiv.org/abs/1808.02990}{{\tt
  arXiv:1808.02990}}].

\bibitem{Ciezarek:2017yzh}
G.~Ciezarek, M.~Franco~Sevilla, B.~Hamilton, R.~Kowalewski, T.~Kuhr, V.~Lüth,
  and Y.~Sato, {\it {A Challenge to Lepton Universality in B Meson Decays}},
  {\it Nature} {\bf 546} (2017) 227--233,
  [\href{http://arxiv.org/abs/1703.01766}{{\tt arXiv:1703.01766}}].

\bibitem{Lees:2012xj}
{\bf BaBar} Collaboration, J.~P. Lees et~al., {\it {Evidence for an excess of
  $\bar{B} \to D^{(*)} \tau^-\bar{\nu}_\tau$ decays}},  {\it Phys. Rev. Lett.}
  {\bf 109} (2012) 101802, [\href{http://arxiv.org/abs/1205.5442}{{\tt
  arXiv:1205.5442}}].

\bibitem{Lees:2013uzd}
{\bf BaBar} Collaboration, J.~P. Lees et~al., {\it {Measurement of an Excess of
  $\bar{B} \to D^{(*)}\tau^- \bar{\nu}_\tau$ Decays and Implications for
  Charged Higgs Bosons}},  {\it Phys. Rev.} {\bf D88} (2013), no.~7 072012,
  [\href{http://arxiv.org/abs/1303.0571}{{\tt arXiv:1303.0571}}].

\bibitem{Huschle:2015rga}
{\bf Belle} Collaboration, M.~Huschle et~al., {\it {Measurement of the
  branching ratio of $\bar{B} \to D^{(\ast)} \tau^- \bar{\nu}_\tau$ relative to
  $\bar{B} \to D^{(\ast)} \ell^- \bar{\nu}_\ell$ decays with hadronic tagging
  at Belle}},  {\it Phys. Rev.} {\bf D92} (2015), no.~7 072014,
  [\href{http://arxiv.org/abs/1507.03233}{{\tt arXiv:1507.03233}}].

\bibitem{Sato:2016svk}
{\bf Belle} Collaboration, Y.~Sato et~al., {\it {Measurement of the branching
  ratio of $\bar{B}^0 \rightarrow D^{*+} \tau^- \bar{\nu}_{\tau}$ relative to
  $\bar{B}^0 \rightarrow D^{*+} \ell^- \bar{\nu}_{\ell}$ decays with a
  semileptonic tagging method}},  {\it Phys. Rev.} {\bf D94} (2016), no.~7
  072007, [\href{http://arxiv.org/abs/1607.07923}{{\tt arXiv:1607.07923}}].

\bibitem{Hirose:2016wfn}
{\bf Belle} Collaboration, S.~Hirose et~al., {\it {Measurement of the $\tau$
  lepton polarization and $R(D^*)$ in the decay $\bar{B} \to D^* \tau^-
  \bar{\nu}_\tau$}},  {\it Phys. Rev. Lett.} {\bf 118} (2017), no.~21 211801,
  [\href{http://arxiv.org/abs/1612.00529}{{\tt arXiv:1612.00529}}].

\bibitem{Hirose:2017dxl}
{\bf Belle} Collaboration, S.~Hirose et~al., {\it {Measurement of the $\tau$
  lepton polarization and $R(D^*)$ in the decay $\bar{B} \rightarrow D^* \tau^-
  \bar{\nu}_\tau$ with one-prong hadronic $\tau$ decays at Belle}},  {\it Phys.
  Rev.} {\bf D97} (2018), no.~1 012004,
  [\href{http://arxiv.org/abs/1709.00129}{{\tt arXiv:1709.00129}}].

\bibitem{Aaij:2015yra}
{\bf LHCb} Collaboration, R.~Aaij et~al., {\it {Measurement of the ratio of
  branching fractions $\mathcal{B}(\bar{B}^0 \to
  D^{*+}\tau^{-}\bar{\nu}_{\tau})/\mathcal{B}(\bar{B}^0 \to
  D^{*+}\mu^{-}\bar{\nu}_{\mu})$}},  {\it Phys. Rev. Lett.} {\bf 115} (2015),
  no.~11 111803, [\href{http://arxiv.org/abs/1506.08614}{{\tt
  arXiv:1506.08614}}]. [Erratum: Phys. Rev. Lett.115,no.15,159901(2015)].

\bibitem{Aaij:2017uff}
{\bf LHCb} Collaboration, R.~Aaij et~al., {\it {Measurement of the ratio of the
  $B^0 \to D^{*-} \tau^+ \nu_{\tau}$ and $B^0 \to D^{*-} \mu^+ \nu_{\mu}$
  branching fractions using three-prong $\tau$-lepton decays}},  {\it Phys.
  Rev. Lett.} {\bf 120} (2018), no.~17 171802,
  [\href{http://arxiv.org/abs/1708.08856}{{\tt arXiv:1708.08856}}].

\bibitem{Aaij:2017deq}
{\bf LHCb} Collaboration, R.~Aaij et~al., {\it {Test of Lepton Flavor
  Universality by the measurement of the $B^0 \to D^{*-} \tau^+ \nu_{\tau}$
  branching fraction using three-prong $\tau$ decays}},  {\it Phys. Rev.} {\bf
  D97} (2018), no.~7 072013, [\href{http://arxiv.org/abs/1711.02505}{{\tt
  arXiv:1711.02505}}].

\bibitem{Amhis:2016xyh}
{\bf HFLAV} Collaboration, Y.~Amhis et~al., {\it {Averages of $b$-hadron,
  $c$-hadron, and $\tau$-lepton properties as of summer 2016}},  {\it Eur.
  Phys. J.} {\bf C77} (2017), no.~12 895,
  [\href{http://arxiv.org/abs/1612.07233}{{\tt arXiv:1612.07233}}].

\bibitem{Amhis:2018up}
{\bf HFLAV} Collaboration, {\it {Online update for averages of $R_D$ and
  $R_{D^{\ast}}$ for Summer 2018 at
  \url{https://hflav-eos.web.cern.ch/hflav-eos/semi/summer18/RDRDs.html}}}, .

\bibitem{Bigi:2016mdz}
D.~Bigi and P.~Gambino, {\it {Revisiting $B\to D \ell \nu$}},  {\it Phys. Rev.}
  {\bf D94} (2016), no.~9 094008, [\href{http://arxiv.org/abs/1606.08030}{{\tt
  arXiv:1606.08030}}].

\bibitem{Bernlochner:2017jka}
F.~U. Bernlochner, Z.~Ligeti, M.~Papucci, and D.~J. Robinson, {\it {Combined
  analysis of semileptonic $B$ decays to $D$ and $D^*$: $R(D^{(*)})$,
  $|V_{cb}|$, and new physics}},  {\it Phys. Rev.} {\bf D95} (2017), no.~11
  115008, [\href{http://arxiv.org/abs/1703.05330}{{\tt arXiv:1703.05330}}].
  [Erratum: Phys. Rev.D97,no.5,059902(2018)].

\bibitem{Bigi:2017jbd}
D.~Bigi, P.~Gambino, and S.~Schacht, {\it {$R(D^*)$, $|V_{cb}|$, and the Heavy
  Quark Symmetry relations between form factors}},  {\it JHEP} {\bf 11} (2017)
  061, [\href{http://arxiv.org/abs/1707.09509}{{\tt arXiv:1707.09509}}].

\bibitem{Jaiswal:2017rve}
S.~Jaiswal, S.~Nandi, and S.~K. Patra, {\it {Extraction of $|V_{cb}|$ from
  $B\to D^{(*)}\ell\nu_\ell$ and the Standard Model predictions of
  $R(D^{(*)})$}},  {\it JHEP} {\bf 12} (2017) 060,
  [\href{http://arxiv.org/abs/1707.09977}{{\tt arXiv:1707.09977}}].

\bibitem{Hu:2018lmk}
Q.-Y. Hu, X.-Q. Li, Y.~Muramatsu, and Y.-D. Yang, {\it {R-parity violating
  solutions to the $R_{D^{(\ast)}}$ anomaly and their GUT-scale unifications}},
   {\it Phys. Rev.} {\bf D99} (2019), no.~1 015008,
  [\href{http://arxiv.org/abs/1808.01419}{{\tt arXiv:1808.01419}}].

\bibitem{Carena:2018cow}
M.~Carena, E.~Megías, M.~Quíros, and C.~Wagner, {\it {$
  {R}_{D^{\left(*\right)}} $ in custodial warped space}},  {\it JHEP} {\bf 12}
  (2018) 043, [\href{http://arxiv.org/abs/1809.01107}{{\tt arXiv:1809.01107}}].

\bibitem{Cohen:2018vhw}
T.~D. Cohen, H.~Lamm, and R.~F. Lebed, {\it {Tests of the standard model in
  $B\to D \ell \nu_{\ell}$, $B\to D^\ast \ell \nu_{\ell}$ and $B_c\to J/\psi
  \ell \nu_{\ell}$}},  {\it Phys. Rev.} {\bf D98} (2018), no.~3 034022,
  [\href{http://arxiv.org/abs/1807.00256}{{\tt arXiv:1807.00256}}].

\bibitem{Heeck:2018ntp}
J.~Heeck and D.~Teresi, {\it {Pati-Salam explanations of the B-meson
  anomalies}},  {\it JHEP} {\bf 12} (2018) 103,
  [\href{http://arxiv.org/abs/1808.07492}{{\tt arXiv:1808.07492}}].

\bibitem{Angelescu:2018tyl}
A.~Angelescu, D.~Be?irevi?, D.~A. Faroughy, and O.~Sumensari, {\it {Closing the
  window on single leptoquark solutions to the $B$-physics anomalies}},  {\it
  JHEP} {\bf 10} (2018) 183, [\href{http://arxiv.org/abs/1808.08179}{{\tt
  arXiv:1808.08179}}].

\bibitem{Huang:2018nnq}
Z.-R. Huang, Y.~Li, C.-D. Lu, M.~A. Paracha, and C.~Wang, {\it {Footprints of
  New Physics in $b\to c\tau\nu$ Transitions}},  {\it Phys. Rev.} {\bf D98}
  (2018), no.~9 095018, [\href{http://arxiv.org/abs/1808.03565}{{\tt
  arXiv:1808.03565}}].

\bibitem{Azatov:2018kzb}
A.~Azatov, D.~Barducci, D.~Ghosh, D.~Marzocca, and L.~Ubaldi, {\it {Combined
  explanations of B-physics anomalies: the sterile neutrino solution}},  {\it
  JHEP} {\bf 10} (2018) 092, [\href{http://arxiv.org/abs/1807.10745}{{\tt
  arXiv:1807.10745}}].

\bibitem{Li:2018rax}
S.-P. Li, X.-Q. Li, Y.-D. Yang, and X.~Zhang, {\it {$
  {R}_{D^{\left(*\right)}},{R}_{K^{\left(*\right)}} $ and neutrino mass in the
  2HDM-III with right-handed neutrinos}},  {\it JHEP} {\bf 09} (2018) 149,
  [\href{http://arxiv.org/abs/1807.08530}{{\tt arXiv:1807.08530}}].

\bibitem{Robinson:2018gza}
D.~J. Robinson, B.~Shakya, and J.~Zupan, {\it {Right-handed neutrinos and
  R(D$^{(?)}$)}},  {\it JHEP} {\bf 02} (2019) 119,
  [\href{http://arxiv.org/abs/1807.04753}{{\tt arXiv:1807.04753}}].

\bibitem{Trifinopoulos:2018rna}
S.~Trifinopoulos, {\it {Revisiting R-parity violating interactions as an
  explanation of the B-physics anomalies}},  {\it Eur. Phys. J.} {\bf C78}
  (2018), no.~10 803, [\href{http://arxiv.org/abs/1807.01638}{{\tt
  arXiv:1807.01638}}].

\bibitem{Feruglio:2018fxo}
F.~Feruglio, P.~Paradisi, and O.~Sumensari, {\it {Implications of scalar and
  tensor explanations of $R_{D^{(\ast)}}$}},  {\it JHEP} {\bf 11} (2018) 191,
  [\href{http://arxiv.org/abs/1806.10155}{{\tt arXiv:1806.10155}}].

\bibitem{Kumar:2018kmr}
J.~Kumar, D.~London, and R.~Watanabe, {\it {Combined Explanations of the $b \to
  s \mu^+ \mu^-$ and $b \to c \tau^- {\bar\nu}$ Anomalies: a General Model
  Analysis}},  {\it Phys. Rev.} {\bf D99} (2019), no.~1 015007,
  [\href{http://arxiv.org/abs/1806.07403}{{\tt arXiv:1806.07403}}].

\bibitem{Becirevic:2018afm}
D.~Be\v{c}irevi\'c, I.~Dor\v{s}ner, S.~Fajfer, D.~A. Faroughy, N.~Ko\v{s}nik,
  and O.~Sumensari, {\it {Scalar leptoquarks from grand unified theories to
  accommodate the $B$-physics anomalies}},  {\it Phys. Rev.} {\bf D98} (2018),
  no.~5 055003, [\href{http://arxiv.org/abs/1806.05689}{{\tt
  arXiv:1806.05689}}].

\bibitem{Dhargyal:2018bbc}
L.~Dhargyal and S.~K. Rai, {\it {Implications of a vector-like lepton doublet
  and scalar Leptoquark on $R(D^{(*)})$}},
  \href{http://arxiv.org/abs/1806.01178}{{\tt arXiv:1806.01178}}.

\bibitem{Bhattacharya:2018kig}
S.~Bhattacharya, S.~Nandi, and S.~Kumar~Patra, {\it {$b \to c \tau \nu_{\tau}$
  Decays: A Catalogue to Compare, Constrain, and Correlate New Physics
  Effects}},  \href{http://arxiv.org/abs/1805.08222}{{\tt arXiv:1805.08222}}.

\bibitem{Fraser:2018aqj}
S.~Fraser, C.~Marzo, L.~Marzola, M.~Raidal, and C.~Spethmann, {\it {Towards a
  viable scalar interpretation of $R_{D^{(*)}}$}},  {\it Phys. Rev.} {\bf D98}
  (2018), no.~3 035016, [\href{http://arxiv.org/abs/1805.08189}{{\tt
  arXiv:1805.08189}}].

\bibitem{Martinez:2018ynq}
R.~Martinez, C.~F. Sierra, and G.~Valencia, {\it {Beyond $\mathcal{R}(D^{(*)})$
  with the general type-III 2HDM for $b\to c\tau\nu$}},  {\it Phys. Rev.} {\bf
  D98} (2018), no.~11 115012, [\href{http://arxiv.org/abs/1805.04098}{{\tt
  arXiv:1805.04098}}].

\bibitem{Azatov:2018knx}
A.~Azatov, D.~Bardhan, D.~Ghosh, F.~Sgarlata, and E.~Venturini, {\it {Anatomy
  of $b \to c \tau \nu$ anomalies}},  {\it JHEP} {\bf 11} (2018) 187,
  [\href{http://arxiv.org/abs/1805.03209}{{\tt arXiv:1805.03209}}].

\bibitem{Alok:2018uft}
A.~K. Alok, D.~Kumar, S.~Kumbhakar, and S.~Uma~Sankar, {\it {Resolution of
  $R_D$/$R_{D^*}$ puzzle}},  {\it Phys. Lett.} {\bf B784} (2018) 16--20,
  [\href{http://arxiv.org/abs/1804.08078}{{\tt arXiv:1804.08078}}].

\bibitem{Aydemir:2018cbb}
U.~Aydemir, D.~Minic, C.~Sun, and T.~Takeuchi, {\it {$B$-decay anomalies and
  scalar leptoquarks in unified Pati-Salam models from noncommutative
  geometry}},  {\it JHEP} {\bf 09} (2018) 117,
  [\href{http://arxiv.org/abs/1804.05844}{{\tt arXiv:1804.05844}}].

\bibitem{Greljo:2018ogz}
A.~Greljo, D.~J. Robinson, B.~Shakya, and J.~Zupan, {\it {$R(D^{(*)})$ from
  $W'$ and right-handed neutrinos}},  {\it JHEP} {\bf 09} (2018) 169,
  [\href{http://arxiv.org/abs/1804.04642}{{\tt arXiv:1804.04642}}].

\bibitem{Asadi:2018wea}
P.~Asadi, M.~R. Buckley, and D.~Shih, {\it {It's all right(-handed neutrinos):
  a new $W'$ model for the $R_{D^{(*)}}$ anomaly}},  {\it JHEP} {\bf 09} (2018)
  010, [\href{http://arxiv.org/abs/1804.04135}{{\tt arXiv:1804.04135}}].

\bibitem{Marzocca:2018wcf}
D.~Marzocca, {\it {Addressing the B-physics anomalies in a fundamental
  Composite Higgs Model}},  {\it JHEP} {\bf 07} (2018) 121,
  [\href{http://arxiv.org/abs/1803.10972}{{\tt arXiv:1803.10972}}].

\bibitem{Yang:2018pyq}
Z.-J. Yang, S.-M. Zhao, X.-X. Dong, X.-J. Zhan, H.-B. Zhang, and T.-F. Feng,
  {\it {Corrections to $R_{D}$ and $R_{D^{*}}$ in the BLMSSM}},  {\it Chin.
  Phys.} {\bf C42} (2018), no.~11 113104,
  [\href{http://arxiv.org/abs/1802.01769}{{\tt arXiv:1802.01769}}].

\bibitem{Iguro:2018qzf}
S.~Iguro and Y.~Omura, {\it {Status of the semileptonic $B$ decays and muon
  $g-2$ in general 2HDMs with right-handed neutrinos}},  {\it JHEP} {\bf 05}
  (2018) 173, [\href{http://arxiv.org/abs/1802.01732}{{\tt arXiv:1802.01732}}].

\bibitem{Colangelo:2018cnj}
P.~Colangelo and F.~De~Fazio, {\it {Scrutinizing $ \overline{B}\to
  {D}^{\ast}\left(D\pi \right){\ell}^{-}{\overline{\nu}}_{\ell } $ and $
  \overline{B}\to {D}^{\ast}\left(D\gamma
  \right){\ell}^{-}{\overline{\nu}}_{\ell } $ in search of new physics
  footprints}},  {\it JHEP} {\bf 06} (2018) 082,
  [\href{http://arxiv.org/abs/1801.10468}{{\tt arXiv:1801.10468}}].

\bibitem{Biswas:2018jun}
A.~Biswas, D.~K. Ghosh, A.~Shaw, and S.~K. Patra, {\it {$b \to c \ell \nu$
  anomalies in light of extended scalar sectors}},
  \href{http://arxiv.org/abs/1801.03375}{{\tt arXiv:1801.03375}}.

\bibitem{Jung:2018lfu}
M.~Jung and D.~M. Straub, {\it {Constraining new physics in $b\to c\ell\nu$
  transitions}},  \href{http://arxiv.org/abs/1801.01112}{{\tt
  arXiv:1801.01112}}.

\bibitem{Choudhury:2017ijp}
D.~Choudhury, A.~Kundu, R.~Mandal, and R.~Sinha, {\it {$R_{K^{(*)}}$ and
  $R(D^{(*)})$ anomalies resolved with lepton mixing}},  {\it Nucl. Phys.} {\bf
  B933} (2018) 433--453, [\href{http://arxiv.org/abs/1712.01593}{{\tt
  arXiv:1712.01593}}].

\bibitem{He:2017bft}
X.-G. He and G.~Valencia, {\it {Lepton universality violation and right-handed
  currents in $b \to c \tau \nu$}},  {\it Phys. Lett.} {\bf B779} (2018)
  52--57, [\href{http://arxiv.org/abs/1711.09525}{{\tt arXiv:1711.09525}}].

\bibitem{Bardhan:2016uhr}
D.~Bardhan, P.~Byakti, and D.~Ghosh, {\it {A closer look at the R$_{D}$ and
  R$_{D^*}$ anomalies}},  {\it JHEP} {\bf 01} (2017) 125,
  [\href{http://arxiv.org/abs/1610.03038}{{\tt arXiv:1610.03038}}].

\bibitem{Crivellin:2015hha}
A.~Crivellin, J.~Heeck, and P.~Stoffer, {\it {A perturbed lepton-specific
  two-Higgs-doublet model facing experimental hints for physics beyond the
  Standard Model}},  {\it Phys. Rev. Lett.} {\bf 116} (2016), no.~8 081801,
  [\href{http://arxiv.org/abs/1507.07567}{{\tt arXiv:1507.07567}}].

\bibitem{Fajfer:2012vx}
S.~Fajfer, J.~F. Kamenik, and I.~Nisandzic, {\it {On the $B \to D^* \tau \bar
  \nu_{\tau}$ Sensitivity to New Physics}},  {\it Phys. Rev.} {\bf D85} (2012)
  094025, [\href{http://arxiv.org/abs/1203.2654}{{\tt arXiv:1203.2654}}].

\bibitem{Fajfer:2012jt}
S.~Fajfer, J.~F. Kamenik, I.~Nisandzic, and J.~Zupan, {\it {Implications of
  Lepton Flavor Universality Violations in B Decays}},  {\it Phys. Rev. Lett.}
  {\bf 109} (2012) 161801, [\href{http://arxiv.org/abs/1206.1872}{{\tt
  arXiv:1206.1872}}].

\bibitem{Crivellin:2012ye}
A.~Crivellin, C.~Greub, and A.~Kokulu, {\it {Explaining $B\to D\tau\nu$, $B\to
  D^*\tau\nu$ and $B\to \tau\nu$ in a 2HDM of type III}},  {\it Phys. Rev.}
  {\bf D86} (2012) 054014, [\href{http://arxiv.org/abs/1206.2634}{{\tt
  arXiv:1206.2634}}].

\bibitem{Alok:2017qsi}
A.~K. Alok, D.~Kumar, J.~Kumar, S.~Kumbhakar, and S.~U. Sankar, {\it {New
  physics solutions for $R_D$ and $R_{D^*}$}},  {\it JHEP} {\bf 09} (2018) 152,
  [\href{http://arxiv.org/abs/1710.04127}{{\tt arXiv:1710.04127}}].

\bibitem{Buchmuller:1985jz}
W.~Buchmuller and D.~Wyler, {\it {Effective Lagrangian Analysis of New
  Interactions and Flavor Conservation}},  {\it Nucl. Phys.} {\bf B268} (1986)
  621--653.

\bibitem{Grzadkowski:2010es}
B.~Grzadkowski, M.~Iskrzynski, M.~Misiak, and J.~Rosiek, {\it {Dimension-Six
  Terms in the Standard Model Lagrangian}},  {\it JHEP} {\bf 10} (2010) 085,
  [\href{http://arxiv.org/abs/1008.4884}{{\tt arXiv:1008.4884}}].

\bibitem{Weinberg:1980wa}
S.~Weinberg, {\it {Effective Gauge Theories}},  {\it Phys. Lett.} {\bf 91B}
  (1980) 51--55.

\bibitem{Coleman:1969sm}
S.~R. Coleman, J.~Wess, and B.~Zumino, {\it {Structure of phenomenological
  Lagrangians. 1.}},  {\it Phys. Rev.} {\bf 177} (1969) 2239--2247.

\bibitem{Callan:1969sn}
C.~G. Callan, Jr., S.~R. Coleman, J.~Wess, and B.~Zumino, {\it {Structure of
  phenomenological Lagrangians. 2.}},  {\it Phys. Rev.} {\bf 177} (1969)
  2247--2250.

\bibitem{Willenbrock:2014bja}
S.~Willenbrock and C.~Zhang, {\it {Effective Field Theory Beyond the Standard
  Model}},  {\it Ann. Rev. Nucl. Part. Sci.} {\bf 64} (2014) 83--100,
  [\href{http://arxiv.org/abs/1401.0470}{{\tt arXiv:1401.0470}}].

\bibitem{Falkowski:2015fla}
A.~Falkowski, {\it {Effective field theory approach to LHC Higgs data}},  {\it
  Pramana} {\bf 87} (2016), no.~3 39,
  [\href{http://arxiv.org/abs/1505.00046}{{\tt arXiv:1505.00046}}].

\bibitem{Brivio:2017vri}
I.~Brivio and M.~Trott, {\it {The Standard Model as an Effective Field
  Theory}},  {\it Phys. Rept.} {\bf 793} (2019) 1--98,
  [\href{http://arxiv.org/abs/1706.08945}{{\tt arXiv:1706.08945}}].

\bibitem{deFlorian:2016spz}
{\bf LHC Higgs Cross Section Working Group} Collaboration, D.~de~Florian
  et~al., {\it {Handbook of LHC Higgs Cross Sections: 4. Deciphering the Nature
  of the Higgs Sector}},  \href{http://arxiv.org/abs/1610.07922}{{\tt
  arXiv:1610.07922}}.

\bibitem{Weinberg:1979sa}
S.~Weinberg, {\it {Baryon and Lepton Nonconserving Processes}},  {\it Phys.
  Rev. Lett.} {\bf 43} (1979) 1566--1570.

\bibitem{Hagiwara:1993ck}
K.~Hagiwara, S.~Ishihara, R.~Szalapski, and D.~Zeppenfeld, {\it {Low-energy
  effects of new interactions in the electroweak boson sector}},  {\it Phys.
  Rev.} {\bf D48} (1993) 2182--2203.

\bibitem{Contino:2013kra}
R.~Contino, M.~Ghezzi, C.~Grojean, M.~Muhlleitner, and M.~Spira, {\it
  {Effective Lagrangian for a light Higgs-like scalar}},  {\it JHEP} {\bf 07}
  (2013) 035, [\href{http://arxiv.org/abs/1303.3876}{{\tt arXiv:1303.3876}}].

\bibitem{Giudice:2007fh}
G.~F. Giudice, C.~Grojean, A.~Pomarol, and R.~Rattazzi, {\it {The
  Strongly-Interacting Light Higgs}},  {\it JHEP} {\bf 06} (2007) 045,
  [\href{http://arxiv.org/abs/hep-ph/0703164}{{\tt hep-ph/0703164}}].

\bibitem{Falkowski:2015wza}
A.~Falkowski, B.~Fuks, K.~Mawatari, K.~Mimasu, F.~Riva, and V.~Sanz, {\it
  {Rosetta: an operator basis translator for Standard Model effective field
  theory}},  {\it Eur. Phys. J.} {\bf C75} (2015), no.~12 583,
  [\href{http://arxiv.org/abs/1508.05895}{{\tt arXiv:1508.05895}}].

\bibitem{Aebischer:2017ugx}
J.~Aebischer et~al., {\it {WCxf: an exchange format for Wilson coefficients
  beyond the Standard Model}},  {\it Comput. Phys. Commun.} {\bf 232} (2018)
  71--83, [\href{http://arxiv.org/abs/1712.05298}{{\tt arXiv:1712.05298}}].

\bibitem{Jenkins:2013zja}
E.~E. Jenkins, A.~V. Manohar, and M.~Trott, {\it {Renormalization Group
  Evolution of the Standard Model Dimension Six Operators I: Formalism and
  lambda Dependence}},  {\it JHEP} {\bf 10} (2013) 087,
  [\href{http://arxiv.org/abs/1308.2627}{{\tt arXiv:1308.2627}}].

\bibitem{Jenkins:2013wua}
E.~E. Jenkins, A.~V. Manohar, and M.~Trott, {\it {Renormalization Group
  Evolution of the Standard Model Dimension Six Operators II: Yukawa
  Dependence}},  {\it JHEP} {\bf 01} (2014) 035,
  [\href{http://arxiv.org/abs/1310.4838}{{\tt arXiv:1310.4838}}].

\bibitem{Alonso:2013hga}
R.~Alonso, E.~E. Jenkins, A.~V. Manohar, and M.~Trott, {\it {Renormalization
  Group Evolution of the Standard Model Dimension Six Operators III: Gauge
  Coupling Dependence and Phenomenology}},  {\it JHEP} {\bf 04} (2014) 159,
  [\href{http://arxiv.org/abs/1312.2014}{{\tt arXiv:1312.2014}}].

\bibitem{Buchalla:1995vs}
G.~Buchalla, A.~J. Buras, and M.~E. Lautenbacher, {\it {Weak decays beyond
  leading logarithms}},  {\it Rev. Mod. Phys.} {\bf 68} (1996) 1125--1144,
  [\href{http://arxiv.org/abs/hep-ph/9512380}{{\tt hep-ph/9512380}}].

\bibitem{Buras:1998raa}
A.~J. Buras, {\it {Weak Hamiltonian, CP violation and rare decays}},  in {\it
  {Probing the standard model of particle interactions. Proceedings, Summer
  School in Theoretical Physics, NATO Advanced Study Institute, 68th session,
  Les Houches, France, July 28-September 5, 1997. Pt. 1, 2}}, pp.~281--539,
  1998.
\newblock \href{http://arxiv.org/abs/hep-ph/9806471}{{\tt hep-ph/9806471}}.

\bibitem{Aebischer:2017gaw}
J.~Aebischer, M.~Fael, C.~Greub, and J.~Virto, {\it {B physics Beyond the
  Standard Model at One Loop: Complete Renormalization Group Evolution below
  the Electroweak Scale}},  {\it JHEP} {\bf 09} (2017) 158,
  [\href{http://arxiv.org/abs/1704.06639}{{\tt arXiv:1704.06639}}].

\bibitem{Jenkins:2017jig}
E.~E. Jenkins, A.~V. Manohar, and P.~Stoffer, {\it {Low-Energy Effective Field
  Theory below the Electroweak Scale: Operators and Matching}},  {\it JHEP}
  {\bf 03} (2018) 016, [\href{http://arxiv.org/abs/1709.04486}{{\tt
  arXiv:1709.04486}}].

\bibitem{Jenkins:2017dyc}
E.~E. Jenkins, A.~V. Manohar, and P.~Stoffer, {\it {Low-Energy Effective Field
  Theory below the Electroweak Scale: Anomalous Dimensions}},  {\it JHEP} {\bf
  01} (2018) 084, [\href{http://arxiv.org/abs/1711.05270}{{\tt
  arXiv:1711.05270}}].

\bibitem{Aebischer:2018bkb}
J.~Aebischer, J.~Kumar, and D.~M. Straub, {\it {Wilson: a Python package for
  the running and matching of Wilson coefficients above and below the
  electroweak scale}},  {\it Eur. Phys. J.} {\bf C78} (2018), no.~12 1026,
  [\href{http://arxiv.org/abs/1804.05033}{{\tt arXiv:1804.05033}}].

\bibitem{Aebischer:2015fzz}
J.~Aebischer, A.~Crivellin, M.~Fael, and C.~Greub, {\it {Matching of gauge
  invariant dimension-six operators for $b\to s$ and $b\to c$ transitions}},
  {\it JHEP} {\bf 05} (2016) 037, [\href{http://arxiv.org/abs/1512.02830}{{\tt
  arXiv:1512.02830}}].

\bibitem{Feruglio:2016gvd}
F.~Feruglio, P.~Paradisi, and A.~Pattori, {\it {Revisiting Lepton Flavor
  Universality in B Decays}},  {\it Phys. Rev. Lett.} {\bf 118} (2017), no.~1
  011801, [\href{http://arxiv.org/abs/1606.00524}{{\tt arXiv:1606.00524}}].

\bibitem{Feruglio:2017rjo}
F.~Feruglio, P.~Paradisi, and A.~Pattori, {\it {On the Importance of
  Electroweak Corrections for B Anomalies}},  {\it JHEP} {\bf 09} (2017) 061,
  [\href{http://arxiv.org/abs/1705.00929}{{\tt arXiv:1705.00929}}].

\bibitem{Gonzalez-Alonso:2017iyc}
M.~Gonz\'alez-Alonso, J.~Martin~Camalich, and K.~Mimouni, {\it
  {Renormalization-group evolution of new physics contributions to
  (semi)leptonic meson decays}},  {\it Phys. Lett.} {\bf B772} (2017) 777--785,
  [\href{http://arxiv.org/abs/1706.00410}{{\tt arXiv:1706.00410}}].

\bibitem{RGEweb}
R.~Alonso, E.~E. Jenkins, A.~V. Manohar, and M.~Trott, {\it {Dimension-Six
  Renormalization Group Equations}},  {\it
  \url{http://einstein.ucsd.edu/smeft/}}.

\bibitem{Dedes:2017zog}
A.~Dedes, W.~Materkowska, M.~Paraskevas, J.~Rosiek, and K.~Suxho, {\it {Feynman
  rules for the Standard Model Effective Field Theory in $R_\xi$-gauges}},
  {\it JHEP} {\bf 06} (2017) 143, [\href{http://arxiv.org/abs/1704.03888}{{\tt
  arXiv:1704.03888}}].

\bibitem{Alonso:2015sja}
R.~Alonso, B.~Grinstein, and J.~Martin~Camalich, {\it {Lepton universality
  violation and lepton flavor conservation in $B$-meson decays}},  {\it JHEP}
  {\bf 10} (2015) 184, [\href{http://arxiv.org/abs/1505.05164}{{\tt
  arXiv:1505.05164}}].

\bibitem{Cata:2015lta}
O.~Cat\'a and M.~Jung, {\it {Signatures of a nonstandard Higgs boson from
  flavor physics}},  {\it Phys. Rev.} {\bf D92} (2015), no.~5 055018,
  [\href{http://arxiv.org/abs/1505.05804}{{\tt arXiv:1505.05804}}].

\bibitem{Cirigliano:2009wk}
V.~Cirigliano, J.~Jenkins, and M.~Gonzalez-Alonso, {\it {Semileptonic decays of
  light quarks beyond the Standard Model}},  {\it Nucl. Phys.} {\bf B830}
  (2010) 95--115, [\href{http://arxiv.org/abs/0908.1754}{{\tt
  arXiv:0908.1754}}].

\bibitem{Capdevila:2017iqn}
B.~Capdevila, A.~Crivellin, S.~Descotes-Genon, L.~Hofer, and J.~Matias, {\it
  {Searching for New Physics with $b\to s\tau^+\tau^-$ processes}},  {\it Phys.
  Rev. Lett.} {\bf 120} (2018), no.~18 181802,
  [\href{http://arxiv.org/abs/1712.01919}{{\tt arXiv:1712.01919}}].

\bibitem{Sirlin:1981ie}
A.~Sirlin, {\it {Large $m_W$, $m_Z$ Behavior of the $O(\alpha)$ Corrections to
  Semileptonic Processes Mediated by W}},  {\it Nucl. Phys.} {\bf B196} (1982)
  83--92.

\bibitem{Marciano:1993sh}
W.~J. Marciano and A.~Sirlin, {\it {Radiative corrections to $\pi_{l2}$
  decays}},  {\it Phys. Rev. Lett.} {\bf 71} (1993) 3629--3632.

\bibitem{Voloshin:1992sn}
M.~B. Voloshin, {\it {Upper bound on tensor interaction in the decay $\pi^- \to
  e^-\nu\gamma$}},  {\it Phys. Lett.} {\bf B283} (1992) 120--122.

\bibitem{Eichten:1989zv}
E.~Eichten and B.~R. Hill, {\it {An Effective Field Theory for the Calculation
  of Matrix Elements Involving Heavy Quarks}},  {\it Phys. Lett.} {\bf B234}
  (1990) 511--516.

\bibitem{Gracey:2000am}
J.~A. Gracey, {\it {Three loop $\overline{\text{MS}}$ tensor current anomalous
  dimension in QCD}},  {\it Phys. Lett.} {\bf B488} (2000) 175--181,
  [\href{http://arxiv.org/abs/hep-ph/0007171}{{\tt hep-ph/0007171}}].

\bibitem{Celis:2017hod}
A.~Celis, J.~Fuentes-Martin, A.~Vicente, and J.~Virto, {\it {DsixTools: The
  Standard Model Effective Field Theory Toolkit}},  {\it Eur. Phys. J.} {\bf
  C77} (2017), no.~6 405, [\href{http://arxiv.org/abs/1704.04504}{{\tt
  arXiv:1704.04504}}].

\bibitem{Celis:2012dk}
A.~Celis, M.~Jung, X.-Q. Li, and A.~Pich, {\it {Sensitivity to charged scalars
  in $B\to D^{(*)}\tau\nu_\tau$ and $B\to\tau\nu_\tau$ decays}},  {\it JHEP}
  {\bf 01} (2013) 054, [\href{http://arxiv.org/abs/1210.8443}{{\tt
  arXiv:1210.8443}}].

\bibitem{Datta:2012qk}
A.~Datta, M.~Duraisamy, and D.~Ghosh, {\it {Diagnosing New Physics in $b \to c
  \, \tau \, \nu_\tau$ decays in the light of the recent BaBar result}},  {\it
  Phys. Rev.} {\bf D86} (2012) 034027,
  [\href{http://arxiv.org/abs/1206.3760}{{\tt arXiv:1206.3760}}].

\bibitem{Sakaki:2013bfa}
Y.~Sakaki, M.~Tanaka, A.~Tayduganov, and R.~Watanabe, {\it {Testing leptoquark
  models in $\bar B \to D^{(*)} \tau \bar\nu$}},  {\it Phys. Rev.} {\bf D88}
  (2013), no.~9 094012, [\href{http://arxiv.org/abs/1309.0301}{{\tt
  arXiv:1309.0301}}].

\bibitem{Gutsche:2015mxa}
T.~Gutsche, M.~A. Ivanov, J.~G. Körner, V.~E. Lyubovitskij, P.~Santorelli, and
  N.~Habyl, {\it {Semileptonic decay $\Lambda_b \to \Lambda_c + \tau^- +
  \bar{\nu_\tau}$ in the covariant confined quark model}},  {\it Phys. Rev.}
  {\bf D91} (2015), no.~7 074001, [\href{http://arxiv.org/abs/1502.04864}{{\tt
  arXiv:1502.04864}}]. [Erratum: Phys. Rev.D91,no.11,119907(2015)].

\bibitem{Li:2016pdv}
X.-Q. Li, Y.-D. Yang, and X.~Zhang, {\it {$ {\varLambda}_b\to
  {\varLambda}_c\tau {\overline{\nu}}_{\tau } $ decay in scalar and vector
  leptoquark scenarios}},  {\it JHEP} {\bf 02} (2017) 068,
  [\href{http://arxiv.org/abs/1611.01635}{{\tt arXiv:1611.01635}}].

\bibitem{Datta:2017aue}
A.~Datta, S.~Kamali, S.~Meinel, and A.~Rashed, {\it {Phenomenology of $
  {\Lambda}_b\to {\Lambda}_c\tau {\overline{\nu}}_{\tau } $ using lattice QCD
  calculations}},  {\it JHEP} {\bf 08} (2017) 131,
  [\href{http://arxiv.org/abs/1702.02243}{{\tt arXiv:1702.02243}}].

\bibitem{Korner:1987kd}
J.~G. Korner and G.~A. Schuler, {\it {Exclusive Semileptonic Decays of Bottom
  Mesons in the Spectator Quark Model}},  {\it Z. Phys.} {\bf C38} (1988) 511.
  [Erratum: Z. Phys.C41,690(1989)].

\bibitem{Korner:1989ve}
J.~G. Korner and G.~A. Schuler, {\it {Lepton Mass Effects in Semileptonic $B$
  Meson Decays}},  {\it Phys. Lett.} {\bf B231} (1989) 306--311.

\bibitem{Korner:1989qb}
J.~G. Korner and G.~A. Schuler, {\it {Exclusive Semileptonic Heavy Meson Decays
  Including Lepton Mass Effects}},  {\it Z. Phys.} {\bf C46} (1990) 93.

\bibitem{Gratrex:2015hna}
J.~Gratrex, M.~Hopfer, and R.~Zwicky, {\it {Generalised helicity formalism,
  higher moments and the $B \to K_{J_K}(\to K \pi) \bar{\ell}_1 \ell_2$ angular
  distributions}},  {\it Phys. Rev.} {\bf D93} (2016), no.~5 054008,
  [\href{http://arxiv.org/abs/1506.03970}{{\tt arXiv:1506.03970}}].

\bibitem{Detmold:2015aaa}
W.~Detmold, C.~Lehner, and S.~Meinel, {\it {$\Lambda_b \to p \ell^-
  \bar{\nu}_\ell$ and $\Lambda_b \to \Lambda_c \ell^- \bar{\nu}_\ell$ form
  factors from lattice QCD with relativistic heavy quarks}},  {\it Phys. Rev.}
  {\bf D92} (2015), no.~3 034503, [\href{http://arxiv.org/abs/1503.01421}{{\tt
  arXiv:1503.01421}}].

\bibitem{Ivanov:2000aj}
M.~A. Ivanov, J.~G. Korner, and P.~Santorelli, {\it {The Semileptonic decays of
  the $B_c$ meson}},  {\it Phys. Rev.} {\bf D63} (2001) 074010,
  [\href{http://arxiv.org/abs/hep-ph/0007169}{{\tt hep-ph/0007169}}].

\bibitem{Ebert:2003cn}
D.~Ebert, R.~N. Faustov, and V.~O. Galkin, {\it {Weak decays of the $B_c$ meson
  to charmonium and $D$ mesons in the relativistic quark model}},  {\it Phys.
  Rev.} {\bf D68} (2003) 094020,
  [\href{http://arxiv.org/abs/hep-ph/0306306}{{\tt hep-ph/0306306}}].

\bibitem{Hernandez:2006gt}
E.~Hernandez, J.~Nieves, and J.~M. Verde-Velasco, {\it {Study of exclusive
  semileptonic and non-leptonic decays of $B_c$ - in a nonrelativistic quark
  model}},  {\it Phys. Rev.} {\bf D74} (2006) 074008,
  [\href{http://arxiv.org/abs/hep-ph/0607150}{{\tt hep-ph/0607150}}].

\bibitem{Ivanov:2006ni}
M.~A. Ivanov, J.~G. K{\"o}rner, and P.~Santorelli, {\it {Exclusive semileptonic
  and nonleptonic decays of the $B_c$ meson}},  {\it Phys. Rev.} {\bf D73}
  (2006) 054024, [\href{http://arxiv.org/abs/hep-ph/0602050}{{\tt
  hep-ph/0602050}}].

\bibitem{Wang:2008xt}
W.~Wang, Y.-L. Shen, and C.-D. Lu, {\it {Covariant Light-Front Approach for
  $B_c$ transition form factors}},  {\it Phys. Rev.} {\bf D79} (2009) 054012,
  [\href{http://arxiv.org/abs/0811.3748}{{\tt arXiv:0811.3748}}].

\bibitem{Qiao:2012vt}
C.-F. Qiao and R.-L. Zhu, {\it {Estimation of semileptonic decays of $B_c$
  meson to S-wave charmonia with nonrelativistic QCD}},  {\it Phys. Rev.} {\bf
  D87} (2013), no.~1 014009, [\href{http://arxiv.org/abs/1208.5916}{{\tt
  arXiv:1208.5916}}].

\bibitem{Wen-Fei:2013uea}
W.-F. Wang, Y.-Y. Fan, and Z.-J. Xiao, {\it {Semileptonic decays
  $B_c\to(\eta_c,J/\Psi)l\nu$ in the perturbative QCD approach}},  {\it Chin.
  Phys.} {\bf C37} (2013) 093102, [\href{http://arxiv.org/abs/1212.5903}{{\tt
  arXiv:1212.5903}}].

\bibitem{Rui:2016opu}
Z.~Rui, H.~Li, G.-x. Wang, and Y.~Xiao, {\it {Semileptonic decays of $B_c$
  meson to S-wave charmonium states in the perturbative QCD approach}},  {\it
  Eur. Phys. J.} {\bf C76} (2016), no.~10 564,
  [\href{http://arxiv.org/abs/1602.08918}{{\tt arXiv:1602.08918}}].

\bibitem{Issadykov:2018myx}
A.~Issadykov and M.~A. Ivanov, {\it {The decays $B_{c}\to
  J/\psi+\bar\ell\nu_\ell$ and $B_{c}\to J/\psi + \pi(K)$ in covariant confined
  quark model}},  {\it Phys. Lett.} {\bf B783} (2018) 178--182,
  [\href{http://arxiv.org/abs/1804.00472}{{\tt arXiv:1804.00472}}].

\bibitem{Kiselev:1999sc}
V.~V. Kiselev, A.~K. Likhoded, and A.~I. Onishchenko, {\it {Semileptonic $B_c$
  meson decays in sum rules of QCD and NRQCD}},  {\it Nucl. Phys.} {\bf B569}
  (2000) 473--504, [\href{http://arxiv.org/abs/hep-ph/9905359}{{\tt
  hep-ph/9905359}}].

\bibitem{Dutta:2017xmj}
R.~Dutta and A.~Bhol, {\it {$B_c \to (J/\psi,\,\eta_c)\tau\nu$ semileptonic
  decays within the standard model and beyond}},  {\it Phys. Rev.} {\bf D96}
  (2017), no.~7 076001, [\href{http://arxiv.org/abs/1701.08598}{{\tt
  arXiv:1701.08598}}].

\bibitem{Watanabe:2017mip}
R.~Watanabe, {\it {New Physics effect on $B_c \to J/\psi \tau\bar\nu$ in
  relation to the $R_{D^{(*)}}$ anomaly}},  {\it Phys. Lett.} {\bf B776} (2018)
  5--9, [\href{http://arxiv.org/abs/1709.08644}{{\tt arXiv:1709.08644}}].

\bibitem{Tran:2018kuv}
C.-T. Tran, M.~A. Ivanov, J.~G. K{\"o}rner, and P.~Santorelli, {\it
  {Implications of new physics in the decays $B_c \to
  (J/\psi,\eta_c)\tau\nu$}},  {\it Phys. Rev.} {\bf D97} (2018), no.~5 054014,
  [\href{http://arxiv.org/abs/1801.06927}{{\tt arXiv:1801.06927}}].

\bibitem{Aaij:2017tyk}
{\bf LHCb} Collaboration, R.~Aaij et~al., {\it {Measurement of the ratio of
  branching fractions
  $\mathcal{B}(B_c^+\,\to\,J/\psi\tau^+\nu_\tau)$/$\mathcal{B}(B_c^+\,\to\,J/\psi\mu^+\nu_\mu)$}},
  {\it Phys. Rev. Lett.} {\bf 120} (2018), no.~12 121801,
  [\href{http://arxiv.org/abs/1711.05623}{{\tt arXiv:1711.05623}}].

\bibitem{Cohen:2018dgz}
T.~D. Cohen, H.~Lamm, and R.~F. Lebed, {\it {Model-independent bounds on
  $R(J/\psi)$}},  {\it JHEP} {\bf 09} (2018) 168,
  [\href{http://arxiv.org/abs/1807.02730}{{\tt arXiv:1807.02730}}].

\bibitem{Murphy:2018sqg}
C.~W. Murphy and A.~Soni, {\it {Model-Independent Determination of $B_c^+ \to
  \eta_c\, \ell^+\, \nu$ Form Factors}},  {\it Phys. Rev.} {\bf D98} (2018),
  no.~9 094026, [\href{http://arxiv.org/abs/1808.05932}{{\tt
  arXiv:1808.05932}}].

\bibitem{Wang:2018duy}
W.~Wang and R.~Zhu, {\it {Model independent investigation of the
  $R_{J/\psi,\eta_c}$ and ratios of decay widths of semileptonic $B_c$ decays
  into a P-wave charmonium}},  \href{http://arxiv.org/abs/1808.10830}{{\tt
  arXiv:1808.10830}}.

\bibitem{Berns:2018vpl}
A.~Berns and H.~Lamm, {\it {Model-Independent Prediction of $R(\eta_c)$}},
  {\it JHEP} {\bf 12} (2018) 114, [\href{http://arxiv.org/abs/1808.07360}{{\tt
  arXiv:1808.07360}}].

\bibitem{Colquhoun:2016osw}
{\bf HPQCD} Collaboration, B.~Colquhoun, C.~Davies, J.~Koponen, A.~Lytle, and
  C.~McNeile, {\it {$B_c$ decays from highly improved staggered quarks and
  NRQCD}},  {\it PoS} {\bf LATTICE2016} (2016) 281,
  [\href{http://arxiv.org/abs/1611.01987}{{\tt arXiv:1611.01987}}].

\bibitem{Lytle:2016ixw}
A.~Lytle, B.~Colquhoun, C.~Davies, J.~Koponen, and C.~McNeile, {\it
  {Semileptonic $B_c$ decays from full lattice QCD}},  {\it PoS} {\bf
  BEAUTY2016} (2016) 069, [\href{http://arxiv.org/abs/1605.05645}{{\tt
  arXiv:1605.05645}}].

\bibitem{Mannel:2017jfk}
T.~Mannel, A.~V. Rusov, and F.~Shahriaran, {\it {Inclusive semitauonic $B$
  decays to order ${\cal O} (\Lambda_{QCD}^3/m_b^3)$}},  {\it Nucl. Phys.} {\bf
  B921} (2017) 211--224, [\href{http://arxiv.org/abs/1702.01089}{{\tt
  arXiv:1702.01089}}].

\bibitem{Falk:1994gw}
A.~F. Falk, Z.~Ligeti, M.~Neubert, and Y.~Nir, {\it {Heavy quark expansion for
  the inclusive decay $\bar{B}\to \tau\bar{\nu}X$}},  {\it Phys. Lett.} {\bf
  B326} (1994) 145--153, [\href{http://arxiv.org/abs/hep-ph/9401226}{{\tt
  hep-ph/9401226}}].

\bibitem{Balk:1993sz}
S.~Balk, J.~G. Korner, D.~Pirjol, and K.~Schilcher, {\it {Inclusive
  semileptonic B decays in QCD including lepton mass effects}},  {\it Z. Phys.}
  {\bf C64} (1994) 37--44, [\href{http://arxiv.org/abs/hep-ph/9312220}{{\tt
  hep-ph/9312220}}].

\bibitem{Czarnecki:1994bn}
A.~Czarnecki, M.~Jezabek, and J.~H. Kuhn, {\it {Radiative corrections to $b\to
  c\tau\bar{\nu}_\tau$}},  {\it Phys. Lett.} {\bf B346} (1995) 335--341,
  [\href{http://arxiv.org/abs/hep-ph/9411282}{{\tt hep-ph/9411282}}].

\bibitem{Jezabek:1996db}
M.~Jezabek and L.~Motyka, {\it {Tau lepton distributions in semileptonic B
  decays}},  {\it Nucl. Phys.} {\bf B501} (1997) 207--223,
  [\href{http://arxiv.org/abs/hep-ph/9701358}{{\tt hep-ph/9701358}}].

\bibitem{Biswas:2009rb}
S.~Biswas and K.~Melnikov, {\it {Second order QCD corrections to inclusive
  semileptonic $b\to X_c l\bar{\nu}_l$ decays with massless and massive
  lepton}},  {\it JHEP} {\bf 02} (2010) 089,
  [\href{http://arxiv.org/abs/0911.4142}{{\tt arXiv:0911.4142}}].

\bibitem{Alberti:2014yda}
A.~Alberti, P.~Gambino, K.~J. Healey, and S.~Nandi, {\it {Precision
  Determination of the Cabibbo-Kobayashi-Maskawa Element $V_{cb}$}},  {\it
  Phys. Rev. Lett.} {\bf 114} (2015), no.~6 061802,
  [\href{http://arxiv.org/abs/1411.6560}{{\tt arXiv:1411.6560}}].

\bibitem{Bigi:1996si}
I.~I.~Y. Bigi, M.~A. Shifman, N.~Uraltsev, and A.~I. Vainshtein, {\it {High
  power $n$ of $m_b$ in beauty widths and $n=5\to \infty$ limit}},  {\it Phys.
  Rev.} {\bf D56} (1997) 4017--4030,
  [\href{http://arxiv.org/abs/hep-ph/9704245}{{\tt hep-ph/9704245}}].

\bibitem{Goldberger:1999yh}
W.~D. Goldberger, {\it {Semileptonic B decays as a probe of new physics}},
  \href{http://arxiv.org/abs/hep-ph/9902311}{{\tt hep-ph/9902311}}.

\bibitem{Grossman:1994ax}
Y.~Grossman and Z.~Ligeti, {\it {The Inclusive $\bar{B}\to\tau\bar{\nu}X$ decay
  in two Higgs doublet models}},  {\it Phys. Lett.} {\bf B332} (1994) 373--380,
  [\href{http://arxiv.org/abs/hep-ph/9403376}{{\tt hep-ph/9403376}}].

\bibitem{Freytsis:2015qca}
M.~Freytsis, Z.~Ligeti, and J.~T. Ruderman, {\it {Flavor models for $\bar{B}
  \to D^{(*)} \tau \bar{\nu}$}},  {\it Phys. Rev.} {\bf D92} (2015), no.~5
  054018, [\href{http://arxiv.org/abs/1506.08896}{{\tt arXiv:1506.08896}}].

\bibitem{Colangelo:2016ymy}
P.~Colangelo and F.~De~Fazio, {\it {Tension in the inclusive versus exclusive
  determinations of $|V_{cb}|$: a possible role of new physics}},  {\it Phys.
  Rev.} {\bf D95} (2017), no.~1 011701,
  [\href{http://arxiv.org/abs/1611.07387}{{\tt arXiv:1611.07387}}].

\bibitem{Li:2016vvp}
X.-Q. Li, Y.-D. Yang, and X.~Zhang, {\it {Revisiting the one leptoquark
  solution to the $R(D^{(\ast)})$ anomalies and its phenomenological
  implications}},  {\it JHEP} {\bf 08} (2016) 054,
  [\href{http://arxiv.org/abs/1605.09308}{{\tt arXiv:1605.09308}}].

\bibitem{Celis:2016azn}
A.~Celis, M.~Jung, X.-Q. Li, and A.~Pich, {\it {Scalar contributions to $b\to c
  (u) \tau \nu$ transitions}},  {\it Phys. Lett.} {\bf B771} (2017) 168--179,
  [\href{http://arxiv.org/abs/1612.07757}{{\tt arXiv:1612.07757}}].

\bibitem{Kamali:2018fhr}
S.~Kamali, A.~Rashed, and A.~Datta, {\it {New physics in inclusive $B \to
  X_c\ell \bar{\nu}$ decay in light of $R(D^{(*)})$ measurements}},  {\it Phys.
  Rev.} {\bf D97} (2018), no.~9 095034,
  [\href{http://arxiv.org/abs/1801.08259}{{\tt arXiv:1801.08259}}].

\bibitem{Gouz:2002kk}
I.~P. Gouz, V.~V. Kiselev, A.~K. Likhoded, V.~I. Romanovsky, and O.~P.
  Yushchenko, {\it {Prospects for the $B_c$ studies at LHCb}},  {\it Phys.
  Atom. Nucl.} {\bf 67} (2004) 1559--1570,
  [\href{http://arxiv.org/abs/hep-ph/0211432}{{\tt hep-ph/0211432}}]. [Yad.
  Fiz.67,1581(2004)].

\bibitem{Alonso:2016oyd}
R.~Alonso, B.~Grinstein, and J.~Martin~Camalich, {\it {Lifetime of $B_c^-$
  Constrains Explanations for Anomalies in $B\to D^{(\ast)}\tau\nu$}},  {\it
  Phys. Rev. Lett.} {\bf 118} (2017), no.~8 081802,
  [\href{http://arxiv.org/abs/1611.06676}{{\tt arXiv:1611.06676}}].

\bibitem{Akeroyd:2017mhr}
A.~G. Akeroyd and C.-H. Chen, {\it {Constraint on the branching ratio of $B_c
  \to \tau \bar{\nu}$ from LEP1 and consequences for $R(D^{(*)})$ anomaly}},
  {\it Phys. Rev.} {\bf D96} (2017), no.~7 075011,
  [\href{http://arxiv.org/abs/1708.04072}{{\tt arXiv:1708.04072}}].

\bibitem{Colquhoun:2015oha}
{\bf HPQCD} Collaboration, B.~Colquhoun, C.~T.~H. Davies, R.~J. Dowdall,
  J.~Kettle, J.~Koponen, G.~P. Lepage, and A.~T. Lytle, {\it {B-meson decay
  constants: a more complete picture from full lattice QCD}},  {\it Phys. Rev.}
  {\bf D91} (2015), no.~11 114509, [\href{http://arxiv.org/abs/1503.05762}{{\tt
  arXiv:1503.05762}}].

\bibitem{Koppenburg:2017mad}
S.~Descotes-Genon and P.~Koppenburg, {\it {The CKM Parameters}},  {\it Ann.
  Rev. Nucl. Part. Sci.} {\bf 67} (2017) 97--127,
  [\href{http://arxiv.org/abs/1702.08834}{{\tt arXiv:1702.08834}}].

\bibitem{Tanabashi:2018oca}
M.~Tanabashi et~al., {\it {Review of Particle Physics}},  {\it Phys. Rev.} {\bf
  D98} (2018), no.~3 030001.

\bibitem{Abbaneo:2001bv}
{\bf CDF, DELPHI, ALEPH, SLD, OPAL, L3} Collaboration, D.~Abbaneo et~al., {\it
  {Combined results on $b$ hadron production rates and decay properties}},
  \href{http://arxiv.org/abs/hep-ex/0112028}{{\tt hep-ex/0112028}}.

\bibitem{Kou:2018nap}
{\bf Belle II} Collaboration, E.~Kou et~al., {\it {The Belle II Physics Book}},
   \href{http://arxiv.org/abs/1808.10567}{{\tt arXiv:1808.10567}}.

\bibitem{Gonzalez-Alonso:2016etj}
M.~Gonz\'alez-Alonso and J.~Martin~Camalich, {\it {Global
  Effective-Field-Theory analysis of New-Physics effects in (semi)leptonic kaon
  decays}},  {\it JHEP} {\bf 12} (2016) 052,
  [\href{http://arxiv.org/abs/1605.07114}{{\tt arXiv:1605.07114}}].

\bibitem{Lutz:2013ftz}
{\bf Belle} Collaboration, O.~Lutz et~al., {\it {Search for $B \to h^{(*)} \nu
  \bar{\nu}$ with the full Belle $\Upsilon(4S)$ data sample}},  {\it Phys.
  Rev.} {\bf D87} (2013), no.~11 111103,
  [\href{http://arxiv.org/abs/1303.3719}{{\tt arXiv:1303.3719}}].

\bibitem{Lees:2013kla}
{\bf BaBar} Collaboration, J.~P. Lees et~al., {\it {Search for $B \to K^{(*)}
  \nu \overline \nu$ and invisible quarkonium decays}},  {\it Phys. Rev.} {\bf
  D87} (2013), no.~11 112005, [\href{http://arxiv.org/abs/1303.7465}{{\tt
  arXiv:1303.7465}}].

\bibitem{Altmannshofer:2009ma}
W.~Altmannshofer, A.~J. Buras, D.~M. Straub, and M.~Wick, {\it {New strategies
  for New Physics search in $B \to K^{*} \nu \bar{\nu}$, $B \to K \nu
  \bar{\nu}$ and $B \to X_{s} \nu \bar{\nu}$ decays}},  {\it JHEP} {\bf 04}
  (2009) 022, [\href{http://arxiv.org/abs/0902.0160}{{\tt arXiv:0902.0160}}].

\bibitem{Bartsch:2009qp}
M.~Bartsch, M.~Beylich, G.~Buchalla, and D.~N. Gao, {\it {Precision Flavour
  Physics with $B \to K \nu \bar\nu$ and $B \to K l^+ l^-$}},  {\it JHEP} {\bf
  11} (2009) 011, [\href{http://arxiv.org/abs/0909.1512}{{\tt
  arXiv:0909.1512}}].

\bibitem{Buras:2014fpa}
A.~J. Buras, J.~Girrbach-Noe, C.~Niehoff, and D.~M. Straub, {\it {$ B\to
  {K}^{\left(\ast \right)}\nu \overline{\nu} $ decays in the Standard Model and
  beyond}},  {\it JHEP} {\bf 02} (2015) 184,
  [\href{http://arxiv.org/abs/1409.4557}{{\tt arXiv:1409.4557}}].

\bibitem{Becirevic:2015asa}
D.~Bečirević, S.~Fajfer, and N.~Košnik, {\it {Lepton flavor nonuniversality
  in $b\to s \ell^+ \ell^-$ processes}},  {\it Phys. Rev.} {\bf D92} (2015),
  no.~1 014016, [\href{http://arxiv.org/abs/1503.09024}{{\tt
  arXiv:1503.09024}}].

\bibitem{Celis:2015ara}
A.~Celis, J.~Fuentes-Martin, M.~Jung, and H.~Serodio, {\it {Family nonuniversal
  $Z\prime$ models with protected flavor-changing interactions}},  {\it Phys.
  Rev.} {\bf D92} (2015), no.~1 015007,
  [\href{http://arxiv.org/abs/1505.03079}{{\tt arXiv:1505.03079}}].

\bibitem{Calibbi:2015kma}
L.~Calibbi, A.~Crivellin, and T.~Ota, {\it {Effective Field Theory Approach to
  $b\to s\ell\ell^{(\prime)}$, $B\to K^{(\ast)}\nu\bar\nu$ and $B\to
  D^{(\ast)}\tau\nu$ with Third Generation Couplings}},  {\it Phys. Rev. Lett.}
  {\bf 115} (2015) 181801, [\href{http://arxiv.org/abs/1506.02661}{{\tt
  arXiv:1506.02661}}].

\bibitem{Bauer:2015knc}
M.~Bauer and M.~Neubert, {\it {Minimal Leptoquark Explanation for the
  $R_{D^{(\ast)}}$, $R_K$ , and $(g-2)_\mu$ Anomalies}},  {\it Phys. Rev.
  Lett.} {\bf 116} (2016), no.~14 141802,
  [\href{http://arxiv.org/abs/1511.01900}{{\tt arXiv:1511.01900}}].

\bibitem{Fajfer:2015ycq}
S.~Fajfer and N.~Ko\v{s}nik, {\it {Vector leptoquark resolution of $R_K$ and
  $R_{D^{(*)}}$ puzzles}},  {\it Phys. Lett.} {\bf B755} (2016) 270--274,
  [\href{http://arxiv.org/abs/1511.06024}{{\tt arXiv:1511.06024}}].

\bibitem{Barbieri:2015yvd}
R.~Barbieri, G.~Isidori, A.~Pattori, and F.~Senia, {\it {Anomalies in
  $B$-decays and $U(2)$ flavour symmetry}},  {\it Eur. Phys. J.} {\bf C76}
  (2016), no.~2 67, [\href{http://arxiv.org/abs/1512.01560}{{\tt
  arXiv:1512.01560}}].

\bibitem{Deshpand:2016cpw}
N.~G. Deshpande and X.-G. He, {\it {Consequences of R-parity violating
  interactions for anomalies in $\bar B\to D^{(\ast)} \tau \bar \nu$ and $b\to
  s \mu^+\mu^-$}},  {\it Eur. Phys. J.} {\bf C77} (2017), no.~2 134,
  [\href{http://arxiv.org/abs/1608.04817}{{\tt arXiv:1608.04817}}].

\bibitem{Altmannshofer:2017poe}
W.~Altmannshofer, P.~Bhupal~Dev, and A.~Soni, {\it {$R_{D^{(*)}}$ anomaly: A
  possible hint for natural supersymmetry with $R$-parity violation}},  {\it
  Phys. Rev.} {\bf D96} (2017), no.~9 095010,
  [\href{http://arxiv.org/abs/1704.06659}{{\tt arXiv:1704.06659}}].

\bibitem{Aad:2015osa}
{\bf ATLAS} Collaboration, G.~Aad et~al., {\it {A search for high-mass
  resonances decaying to $\tau^{+}\tau^{-}$ in $pp$ collisions at $\sqrt{s}=8$
  TeV with the ATLAS detector}},  {\it JHEP} {\bf 07} (2015) 157,
  [\href{http://arxiv.org/abs/1502.07177}{{\tt arXiv:1502.07177}}].

\bibitem{Aaboud:2016cre}
{\bf ATLAS} Collaboration, M.~Aaboud et~al., {\it {Search for Minimal
  Supersymmetric Standard Model Higgs bosons $H/A$ and for a $Z^{\prime}$ boson
  in the $\tau \tau$ final state produced in $pp$ collisions at $\sqrt{s}=13$
  TeV with the ATLAS Detector}},  {\it Eur. Phys. J.} {\bf C76} (2016), no.~11
  585, [\href{http://arxiv.org/abs/1608.00890}{{\tt arXiv:1608.00890}}].

\bibitem{Faroughy:2016osc}
D.~A. Faroughy, A.~Greljo, and J.~F. Kamenik, {\it {Confronting lepton flavor
  universality violation in B decays with high-$p_T$ tau lepton searches at
  LHC}},  {\it Phys. Lett.} {\bf B764} (2017) 126--134,
  [\href{http://arxiv.org/abs/1609.07138}{{\tt arXiv:1609.07138}}].

\bibitem{Adamczyk:2018}
K.~Adamczyk, {\it {B to semitauonic decays at Belle/Belle II, {\rm Talk given
  at CKM2018}}}, .

\bibitem{Feldmann:2011xf}
T.~Feldmann and M.~W.~Y. Yip, {\it {Form Factors for $\Lambda_b \to \Lambda$
  Transitions in {SCET}}},  {\it Phys. Rev.} {\bf D85} (2012) 014035,
  [\href{http://arxiv.org/abs/1111.1844}{{\tt arXiv:1111.1844}}]. [Erratum:
  Phys. Rev.D86,079901(2012)].

\bibitem{Tanaka:2012nw}
M.~Tanaka and R.~Watanabe, {\it {New physics in the weak interaction of $\bar
  B\to D^{(*)}\tau\bar\nu$}},  {\it Phys. Rev.} {\bf D87} (2013), no.~3 034028,
  [\href{http://arxiv.org/abs/1212.1878}{{\tt arXiv:1212.1878}}].

\end{thebibliography}\endgroup

\end{document}